\documentclass[12pt]{article}

\usepackage[margin=1in]{geometry}
\usepackage{amsmath,amssymb,amsthm,mathtools,bm}
\usepackage{booktabs}
\usepackage{array}
\usepackage{float}
\usepackage{tikz}
\usetikzlibrary{arrows.meta,calc}
\usepackage{microtype}
\usepackage[hidelinks]{hyperref}
\usepackage{xcolor}
\newtheorem{theorem}{Theorem}[section]
\newtheorem{lemma}[theorem]{Lemma}

\newcommand{\R}{\mathbb{R}}
\newcommand{\eZ}{\mathbf e_z}
\newcommand{\norm}[1]{\left\lVert #1\right\rVert}
\newcommand{\Bcal}{\mathcal B}
\newcommand{\Ocal}{\mathcal O}
\newcommand{\Tcal}{\mathcal T}
\newcommand{\Scal}{\mathcal S}

\title{Finite Bodies on Norton's Dome:\\
Escape-Time Divergence and a Center-of-Mass Barrier Theorem}

\author{Guan-Chuan Chen \\
\small Taichung Municipal Taichung First \\
\small Senior High School, Taiwan, R.O.C.
\and
En-Jui Kuo\thanks{Corresponding author: \href{mailto:kuoenjui@nycu.edu.tw}{kuoenjui@nycu.edu.tw}.} \\
\small Department of Electrophysics \\
\small National Yang Ming Chiao Tung University \\
\small Hsinchu, Taiwan, R.O.C.
}

\date{}

\begin{document}
\maketitle
\begin{abstract}
Norton's dome illustrates how a Newtonian initial-value problem can admit
both rest and delayed departure from the same initial state. We examine
how this nonuniqueness changes when the point particle is replaced by a
rigid sphere of positive radius. For a generalized family of Norton's
domes, we show that any fixed positive sphere radius prevents spontaneous
departure from rest at the apex. We then extend the analysis to more
general bodies and establish two sufficient conditions: a geometric
constraint on the distance between the center of mass and the apex, and
an interior tangent ball condition for rigid bodies whose center of mass
is vertically aligned with the initial contact point. Both arguments
combine geometric estimates with an energy inequality and Gr\"onwall's
inequality. The first condition also applies to passive deformable bodies
without releasable internal energy. These results identify explicit
conditions under which finite-body geometry excludes the spontaneous
motion allowed by Norton's point-particle model.
\end{abstract}


\section{Introduction}

Norton's dome allows a point particle initially at rest at the apex
to remain there indefinitely or to begin moving after an arbitrary
waiting time \cite{Norton2008}. The same initial position and velocity
therefore do not determine a unique subsequent motion. We investigate
whether this nonuniqueness persists when the point particle is replaced
by a body of finite size.

Related examples of nonunique motion in classical mechanics were
studied by Bhat and Bernstein \cite{BhatBernstein1997}. Subsequent work
has examined the physical interpretation of Norton's dome and the
assumptions underlying its dynamics
\cite{Malament2008,Fletcher2012,Laraudogoitia2013,Fletcher2019,
VanpouckeWenmackers2021}. Berry discussed the dome alongside quantum
and chaotic systems \cite{Berry2025}, while Korolev considered elastic
effects in Norton-type systems \cite{KorolevElastic}. Our analysis
concerns the contact geometry of finite bodies on a rigid support.

We first consider a rigid sphere on the standard Norton dome and
a generalized family of dome profiles. The sphere's center follows
an offset curve, which changes both the distance travelled and the
gravitational height loss near the apex. We show that every fixed
positive sphere radius makes the escape-time integral diverge,
excluding spontaneous departure from exact rest throughout the
family considered. In the regime where the point-particle model
admits delayed departure, the sphere's height loss has a universal
quadratic dependence on its center arclength to leading order.
This geometric effect, rather than rotational inertia, is responsible
for the exclusion of spontaneous departure.

We then establish two sufficient conditions extending the result
to more general bodies. The first requires the center of mass to
remain at least its initial distance from the fixed apex. Combined
with an energy inequality, this condition excludes departure for
passive systems, including deformable bodies without releasable
internal energy. The second applies to rigid bodies containing an
interior ball tangent at the initial contact point, with the center
of mass vertically aligned with that point. Each condition yields
a quadratic bound on gravitational height loss; kinetic-energy
estimates and Gr\"onwall's inequality (Lemma~\ref{lem:gronwall}) then imply that a body starting
from rest remains stationary. The conclusions depend on these
explicit geometric and energetic hypotheses and concern uniqueness
from exact rest, not stability under perturbations.

Sections~\ref{sec:point} and \ref{sec:generalized} develop the sphere
calculations for the standard and generalized domes.
Sections~\ref{sec:com-barrier} and \ref{sec:rigid} establish the
center-of-mass barrier and interior-tangent-ball criteria.
Section~\ref{sec:essence} summarizes their common mechanism and the
progression of the results. Section~\ref{sec:discussion} discusses their scope
and the singular limit of vanishing sphere radius.

\paragraph{A uniqueness estimate used below.}
The general-body proofs use the following constant-coefficient form of
Gr\"onwall's inequality \cite{Gronwall1919}. It turns a bound on the growth
of a nonnegative displacement into a uniqueness statement. We will discuss its impact on the summary section.

\begin{lemma}[Gr\"onwall's inequality]
\label{lem:gronwall}
Let $D:[0,T]\to[0,\infty)$ be continuous. If constants $A,L\ge0$ satisfy
\begin{equation}
 D(t)\le A+L\int_0^t D(s)\,ds,
\end{equation}
then $D(t)\le A e^{Lt}$. In particular, $A=0$ implies $D\equiv0$.
Consequently, if $D$ is absolutely continuous, $D(0)=0$, and
$D'(t)\le LD(t)$ almost everywhere, then $D\equiv0$.
\end{lemma}

\begin{proof}
Set $F(t)=A+L\int_0^t D(s)\,ds$. Then $D\le F$ and
$F'=LD\le LF$, so $(e^{-Lt}F(t))'\le0$. Since $F(0)=A$,
we obtain $D(t)\le F(t)\le Ae^{Lt}$.
For the differential form, integrate $D'\le LD$ and use $D(0)=0$
to obtain the integral inequality with $A=0$.
\end{proof}

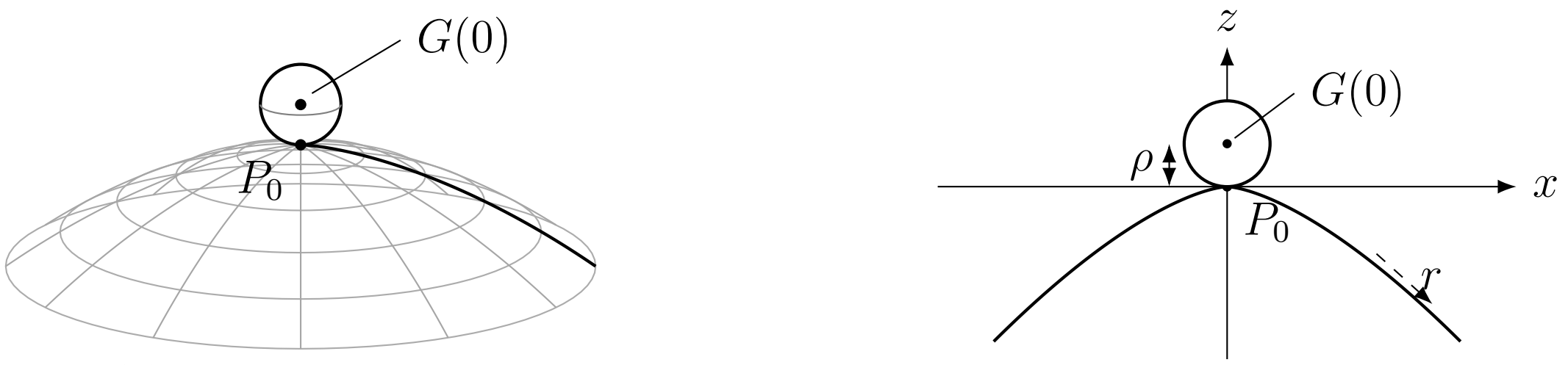
\begin{figure}[H]
\centering
\begin{minipage}[b]{0.48\textwidth}
\centering
\begin{tikzpicture}[scale=1.45,>=Latex]
  \path[use as bounding box] (-2.05,-1.65) rectangle (2.05,1.00);
  \draw[gray!65,thin] plot coordinates {
    (0.3935300,-0.0674619) (0.3923169,-0.0588166) (0.3886850,-0.0502247) (0.3826568,-0.0417389) (0.3742693,-0.0334118)
    (0.3635743,-0.0252946) (0.3506378,-0.0174374) (0.3355395,-0.0098886) (0.3183725,-0.0026948) (0.2992426,0.0040997)
    (0.2782678,0.0104530) (0.2555773,0.0163260) (0.2313112,0.0216824) (0.2056189,0.0264891) (0.1786589,0.0307167)
    (0.1505974,0.0343389) (0.1216075,0.0373335) (0.0918678,0.0396820) (0.0615617,0.0413699) (0.0308760,0.0423868)
    (0.0000000,0.0427265) (-0.0308760,0.0423868) (-0.0615617,0.0413699) (-0.0918678,0.0396820) (-0.1216075,0.0373335)
    (-0.1505974,0.0343389) (-0.1786589,0.0307167) (-0.2056189,0.0264891) (-0.2313112,0.0216824) (-0.2555773,0.0163260)
    (-0.2782678,0.0104530) (-0.2992426,0.0040997) (-0.3183725,-0.0026948) (-0.3355395,-0.0098886) (-0.3506378,-0.0174374)
    (-0.3635743,-0.0252946) (-0.3742693,-0.0334118) (-0.3826568,-0.0417389) (-0.3886850,-0.0502247) (-0.3923169,-0.0588166)
    (-0.3935300,-0.0674619) (-0.3923169,-0.0761072) (-0.3886850,-0.0846992) (-0.3826568,-0.0931849) (-0.3742693,-0.1015120)
    (-0.3635743,-0.1096292) (-0.3506378,-0.1174864) (-0.3355395,-0.1250352) (-0.3183725,-0.1322290) (-0.2992426,-0.1390236)
    (-0.2782678,-0.1453769) (-0.2555773,-0.1512498) (-0.2313112,-0.1566062) (-0.2056189,-0.1614130) (-0.1786589,-0.1656405)
    (-0.1505974,-0.1692627) (-0.1216075,-0.1722573) (-0.0918678,-0.1746058) (-0.0615617,-0.1762937) (-0.0308760,-0.1773107)
    (-0.0000000,-0.1776503) (0.0308760,-0.1773107) (0.0615617,-0.1762937) (0.0918678,-0.1746058) (0.1216075,-0.1722573)
    (0.1505974,-0.1692627) (0.1786589,-0.1656405) (0.2056189,-0.1614130) (0.2313112,-0.1566062) (0.2555773,-0.1512498)
    (0.2782678,-0.1453769) (0.2992426,-0.1390236) (0.3183725,-0.1322290) (0.3355395,-0.1250352) (0.3506378,-0.1174864)
    (0.3635743,-0.1096292) (0.3742693,-0.1015120) (0.3826568,-0.0931849) (0.3886850,-0.0846992) (0.3923169,-0.0761072)
    (0.3935300,-0.0674619)
  };
  \draw[gray!65,thin] plot coordinates {
    (0.7738258,-0.1908111) (0.7714404,-0.1738113) (0.7642988,-0.1569163) (0.7524450,-0.1402302) (0.7359521,-0.1238560)
    (0.7149219,-0.1078946) (0.6894839,-0.0924445) (0.6597950,-0.0776007) (0.6260383,-0.0634550) (0.5884218,-0.0500944)
    (0.5471775,-0.0376014) (0.5025597,-0.0260530) (0.4548434,-0.0155204) (0.4043229,-0.0060685) (0.3513096,0.0022443)
    (0.2961303,0.0093670) (0.2391253,0.0152555) (0.1806461,0.0198735) (0.1210530,0.0231925) (0.0607137,0.0251922)
    (0.0000000,0.0258601) (-0.0607137,0.0251922) (-0.1210530,0.0231925) (-0.1806461,0.0198735) (-0.2391253,0.0152555)
    (-0.2961303,0.0093670) (-0.3513096,0.0022443) (-0.4043229,-0.0060685) (-0.4548434,-0.0155204) (-0.5025597,-0.0260530)
    (-0.5471775,-0.0376014) (-0.5884218,-0.0500944) (-0.6260383,-0.0634550) (-0.6597950,-0.0776007) (-0.6894839,-0.0924445)
    (-0.7149219,-0.1078946) (-0.7359521,-0.1238560) (-0.7524450,-0.1402302) (-0.7642988,-0.1569163) (-0.7714404,-0.1738113)
    (-0.7738258,-0.1908111) (-0.7714404,-0.2078110) (-0.7642988,-0.2247060) (-0.7524450,-0.2413920) (-0.7359521,-0.2577662)
    (-0.7149219,-0.2737276) (-0.6894839,-0.2891778) (-0.6597950,-0.3040215) (-0.6260383,-0.3181673) (-0.5884218,-0.3315278)
    (-0.5471775,-0.3440208) (-0.5025597,-0.3555692) (-0.4548434,-0.3661018) (-0.4043229,-0.3755537) (-0.3513096,-0.3838666)
    (-0.2961303,-0.3909893) (-0.2391253,-0.3968777) (-0.1806461,-0.4014957) (-0.1210530,-0.4048148) (-0.0607137,-0.4068144)
    (-0.0000000,-0.4074824) (0.0607137,-0.4068144) (0.1210530,-0.4048148) (0.1806461,-0.4014957) (0.2391253,-0.3968777)
    (0.2961303,-0.3909893) (0.3513096,-0.3838666) (0.4043229,-0.3755537) (0.4548434,-0.3661018) (0.5025597,-0.3555692)
    (0.5471775,-0.3440208) (0.5884218,-0.3315278) (0.6260383,-0.3181673) (0.6597950,-0.3040215) (0.6894839,-0.2891778)
    (0.7149219,-0.2737276) (0.7359521,-0.2577662) (0.7524450,-0.2413920) (0.7642988,-0.2247060) (0.7714404,-0.2078110)
    (0.7738258,-0.1908111)
  };
  \draw[gray!65,thin] plot coordinates {
    (1.1404097,-0.3505424) (1.1368942,-0.3254893) (1.1263694,-0.3005906) (1.1089001,-0.2759999) (1.0845941,-0.2518688)
    (1.0536012,-0.2283460) (1.0161125,-0.2055766) (0.9723591,-0.1837010) (0.9226108,-0.1628540) (0.8671744,-0.1431641)
    (0.8063914,-0.1247528) (0.7406369,-0.1077336) (0.6703160,-0.0922114) (0.5958624,-0.0782819) (0.5177352,-0.0660309)
    (0.4364159,-0.0555341) (0.3524060,-0.0468561) (0.2662234,-0.0400504) (0.1783994,-0.0351590) (0.0894755,-0.0322121)
    (0.0000000,-0.0312277) (-0.0894755,-0.0322121) (-0.1783994,-0.0351590) (-0.2662234,-0.0400504) (-0.3524060,-0.0468561)
    (-0.4364159,-0.0555341) (-0.5177352,-0.0660309) (-0.5958624,-0.0782819) (-0.6703160,-0.0922114) (-0.7406369,-0.1077336)
    (-0.8063914,-0.1247528) (-0.8671744,-0.1431641) (-0.9226108,-0.1628540) (-0.9723591,-0.1837010) (-1.0161125,-0.2055766)
    (-1.0536012,-0.2283460) (-1.0845941,-0.2518688) (-1.1089001,-0.2759999) (-1.1263694,-0.3005906) (-1.1368942,-0.3254893)
    (-1.1404097,-0.3505424) (-1.1368942,-0.3755956) (-1.1263694,-0.4004943) (-1.1089001,-0.4250850) (-1.0845941,-0.4492161)
    (-1.0536012,-0.4727389) (-1.0161125,-0.4955083) (-0.9723591,-0.5173839) (-0.9226108,-0.5382309) (-0.8671744,-0.5579208)
    (-0.8063914,-0.5763320) (-0.7406369,-0.5933513) (-0.6703160,-0.6088735) (-0.5958624,-0.6228030) (-0.5177352,-0.6350539)
    (-0.4364159,-0.6455508) (-0.3524060,-0.6542288) (-0.2662234,-0.6610345) (-0.1783994,-0.6659259) (-0.0894755,-0.6688728)
    (-0.0000000,-0.6698572) (0.0894755,-0.6688728) (0.1783994,-0.6659259) (0.2662234,-0.6610345) (0.3524060,-0.6542288)
    (0.4364159,-0.6455508) (0.5177352,-0.6350539) (0.5958624,-0.6228030) (0.6703160,-0.6088735) (0.7406369,-0.5933513)
    (0.8063914,-0.5763320) (0.8671744,-0.5579208) (0.9226108,-0.5382309) (0.9723591,-0.5173839) (1.0161125,-0.4955083)
    (1.0536012,-0.4727389) (1.0845941,-0.4492161) (1.1089001,-0.4250850) (1.1263694,-0.4004943) (1.1368942,-0.3755956)
    (1.1404097,-0.3505424)
  };
  \draw[gray!65,thin] plot coordinates {
    (1.4927482,-0.5396954) (1.4881465,-0.5069019) (1.4743700,-0.4743106) (1.4515034,-0.4421223) (1.4196879,-0.4105357)
    (1.3791195,-0.3797454) (1.3300484,-0.3499412) (1.2727771,-0.3213069) (1.2076587,-0.2940191) (1.1350946,-0.2682459)
    (1.0555324,-0.2441463) (0.9694624,-0.2218689) (0.8774154,-0.2015510) (0.7799588,-0.1833178) (0.6776935,-0.1672818)
    (0.5712500,-0.1535419) (0.4612846,-0.1421828) (0.3484751,-0.1332744) (0.2335173,-0.1268718) (0.1171197,-0.1230144)
    (0.0000000,-0.1217259) (-0.1171197,-0.1230144) (-0.2335173,-0.1268718) (-0.3484751,-0.1332744) (-0.4612846,-0.1421828)
    (-0.5712500,-0.1535419) (-0.6776935,-0.1672818) (-0.7799588,-0.1833178) (-0.8774154,-0.2015510) (-0.9694624,-0.2218689)
    (-1.0555324,-0.2441463) (-1.1350946,-0.2682459) (-1.2076587,-0.2940191) (-1.2727771,-0.3213069) (-1.3300484,-0.3499412)
    (-1.3791195,-0.3797454) (-1.4196879,-0.4105357) (-1.4515034,-0.4421223) (-1.4743700,-0.4743106) (-1.4881465,-0.5069019)
    (-1.4927482,-0.5396954) (-1.4881465,-0.5724889) (-1.4743700,-0.6050802) (-1.4515034,-0.6372684) (-1.4196879,-0.6688551)
    (-1.3791195,-0.6996454) (-1.3300484,-0.7294496) (-1.2727771,-0.7580838) (-1.2076587,-0.7853717) (-1.1350946,-0.8111449)
    (-1.0555324,-0.8352445) (-0.9694624,-0.8575219) (-0.8774154,-0.8778398) (-0.7799588,-0.8960730) (-0.6776935,-0.9121089)
    (-0.5712500,-0.9258488) (-0.4612846,-0.9372080) (-0.3484751,-0.9461164) (-0.2335173,-0.9525190) (-0.1171197,-0.9563764)
    (-0.0000000,-0.9576649) (0.1171197,-0.9563764) (0.2335173,-0.9525190) (0.3484751,-0.9461164) (0.4612846,-0.9372080)
    (0.5712500,-0.9258488) (0.6776935,-0.9121089) (0.7799588,-0.8960730) (0.8774154,-0.8778398) (0.9694624,-0.8575219)
    (1.0555324,-0.8352445) (1.1350946,-0.8111449) (1.2076587,-0.7853717) (1.2727771,-0.7580838) (1.3300484,-0.7294496)
    (1.3791195,-0.6996454) (1.4196879,-0.6688551) (1.4515034,-0.6372684) (1.4743700,-0.6050802) (1.4881465,-0.5724889)
    (1.4927482,-0.5396954)
  };
  \draw[gray!65,thin] plot coordinates {
    (1.8302401,-0.7542472) (1.8245981,-0.7140395) (1.8077069,-0.6740797) (1.7796705,-0.6346141) (1.7406618,-0.5958861)
    (1.6909214,-0.5581345) (1.6307559,-0.5215920) (1.5605363,-0.4864838) (1.4806954,-0.4530265) (1.3917255,-0.4214264)
    (1.2941752,-0.3918782) (1.1886459,-0.3645641) (1.0757882,-0.3396525) (0.9562978,-0.3172971) (0.8309116,-0.2976356)
    (0.7004026,-0.2807892) (0.5655753,-0.2668619) (0.4272611,-0.2559395) (0.2863126,-0.2480893) (0.1435990,-0.2433598)
    (0.0000000,-0.2417800) (-0.1435990,-0.2433598) (-0.2863126,-0.2480893) (-0.4272611,-0.2559395) (-0.5655753,-0.2668619)
    (-0.7004026,-0.2807892) (-0.8309116,-0.2976356) (-0.9562978,-0.3172971) (-1.0757882,-0.3396525) (-1.1886459,-0.3645641)
    (-1.2941752,-0.3918782) (-1.3917255,-0.4214264) (-1.4806954,-0.4530265) (-1.5605363,-0.4864838) (-1.6307559,-0.5215920)
    (-1.6909214,-0.5581345) (-1.7406618,-0.5958861) (-1.7796705,-0.6346141) (-1.8077069,-0.6740797) (-1.8245981,-0.7140395)
    (-1.8302401,-0.7542472) (-1.8245981,-0.7944549) (-1.8077069,-0.8344148) (-1.7796705,-0.8738803) (-1.7406618,-0.9126083)
    (-1.6909214,-0.9503600) (-1.6307559,-0.9869025) (-1.5605363,-1.0220106) (-1.4806954,-1.0554679) (-1.3917255,-1.0870681)
    (-1.2941752,-1.1166163) (-1.1886459,-1.1439304) (-1.0757882,-1.1688419) (-0.9562978,-1.1911974) (-0.8309116,-1.2108589)
    (-0.7004026,-1.2277052) (-0.5655753,-1.2416325) (-0.4272611,-1.2525550) (-0.2863126,-1.2604052) (-0.1435990,-1.2651347)
    (-0.0000000,-1.2667145) (0.1435990,-1.2651347) (0.2863126,-1.2604052) (0.4272611,-1.2525550) (0.5655753,-1.2416325)
    (0.7004026,-1.2277052) (0.8309116,-1.2108589) (0.9562978,-1.1911974) (1.0757882,-1.1688419) (1.1886459,-1.1439304)
    (1.2941752,-1.1166163) (1.3917255,-1.0870681) (1.4806954,-1.0554679) (1.5605363,-1.0220106) (1.6307559,-0.9869025)
    (1.6909214,-0.9503600) (1.7406618,-0.9126083) (1.7796705,-0.8738803) (1.8077069,-0.8344148) (1.8245981,-0.7944549)
    (1.8302401,-0.7542472)
  };
  \draw[gray!70,thin] plot coordinates {
    (0.0000000,0.0000000) (0.0498999,-0.0029814) (0.0995989,-0.0084327) (0.1490964,-0.0154919) (0.1983914,-0.0238514)
    (0.2474831,-0.0333333) (0.2963707,-0.0438178) (0.3450533,-0.0552167) (0.3935300,-0.0674619) (0.4418001,-0.0804984)
    (0.4898625,-0.0942809) (0.5377164,-0.1087709) (0.5853609,-0.1239355) (0.6327950,-0.1397458) (0.6800178,-0.1561765)
    (0.7270284,-0.1732051) (0.7738258,-0.1908111) (0.8204091,-0.2089763) (0.8667771,-0.2276840) (0.9129290,-0.2469188)
    (0.9588637,-0.2666667) (1.0045802,-0.2869146) (1.0500774,-0.3076506) (1.0953542,-0.3288634) (1.1404097,-0.3505424)
    (1.1852427,-0.3726780) (1.2298521,-0.3952608) (1.2742367,-0.4182822) (1.3183955,-0.4417340) (1.3623272,-0.4656083)
    (1.4060307,-0.4898979) (1.4495047,-0.5145959) (1.4927482,-0.5396954) (1.5357598,-0.5651902) (1.5785382,-0.5910744)
    (1.6210823,-0.6173420) (1.6633906,-0.6439876) (1.7054619,-0.6710059) (1.7472948,-0.6983918) (1.7888881,-0.7261405)
    (1.8302401,-0.7542472)
  };
  \draw[gray!70,thin] plot coordinates {
    (0.0000000,0.0000000) (0.0432146,0.0040046) (0.0862552,0.0055111) (0.1291212,0.0053816) (0.1718120,0.0039234)
    (0.2143266,0.0013143) (0.2566645,-0.0023259) (0.2988249,-0.0069093) (0.3408070,-0.0123677) (0.3826101,-0.0186464)
    (0.4242334,-0.0257002) (0.4656761,-0.0334906) (0.5069374,-0.0419849) (0.5480166,-0.0511545) (0.5889127,-0.0609740)
    (0.6296251,-0.0714211) (0.6701528,-0.0824755) (0.7104951,-0.0941191) (0.7506510,-0.1063352) (0.7906197,-0.1191087)
    (0.8304003,-0.1324258) (0.8699919,-0.1462734) (0.9093937,-0.1606398) (0.9486046,-0.1755138) (0.9876238,-0.1908851)
    (1.0264503,-0.2067440) (1.0650831,-0.2230815) (1.1035214,-0.2398891) (1.1417640,-0.2571586) (1.1798099,-0.2748825)
    (1.2176583,-0.2930537) (1.2553079,-0.3116652) (1.2927579,-0.3307106) (1.3300070,-0.3501839) (1.3670542,-0.3700790)
    (1.4038984,-0.3903905) (1.4405385,-0.4111129) (1.4769733,-0.4322412) (1.5132017,-0.4537705) (1.5492225,-0.4756962)
    (1.5850345,-0.4980136)
  };
  \draw[gray!70,thin] plot coordinates {
    (0.0000000,0.0000000) (0.0249499,0.0091187) (0.0497995,0.0157187) (0.0745482,0.0206620) (0.0991957,0.0242560)
    (0.1237415,0.0266781) (0.1481853,0.0280483) (0.1725266,0.0284542) (0.1967650,0.0279640) (0.2209000,0.0266324)
    (0.2449312,0.0245044) (0.2688582,0.0216184) (0.2926804,0.0180070) (0.3163975,0.0136988) (0.3400089,0.0087190)
    (0.3635142,0.0030899) (0.3869129,-0.0031683) (0.4102045,-0.0100377) (0.4333886,-0.0175017) (0.4564645,-0.0255453)
    (0.4794318,-0.0341546) (0.5022901,-0.0433169) (0.5250387,-0.0530204) (0.5476771,-0.0632541) (0.5702049,-0.0740078)
    (0.5926213,-0.0852719) (0.6149260,-0.0970375) (0.6371184,-0.1092962) (0.6591977,-0.1220400) (0.6811636,-0.1352615)
    (0.7030153,-0.1489536) (0.7247524,-0.1631096) (0.7463741,-0.1777232) (0.7678799,-0.1927883) (0.7892691,-0.2082992)
    (0.8105411,-0.2242504) (0.8316953,-0.2406368) (0.8527310,-0.2574533) (0.8736474,-0.2746953) (0.8944440,-0.2923582)
    (0.9151201,-0.3104376)
  };
  \draw[gray!70,thin] plot coordinates {
    (0.0000000,0.0000000) (0.0000000,0.0109905) (0.0000000,0.0194550) (0.0000000,0.0262550) (0.0000000,0.0316982)
    (0.0000000,0.0359619) (0.0000000,0.0391660) (0.0000000,0.0413982) (0.0000000,0.0427265) (0.0000000,0.0432056)
    (0.0000000,0.0428806) (0.0000000,0.0417897) (0.0000000,0.0399656) (0.0000000,0.0374368) (0.0000000,0.0342285)
    (0.0000000,0.0303629) (0.0000000,0.0258601) (0.0000000,0.0207382) (0.0000000,0.0150136) (0.0000000,0.0087013)
    (0.0000000,0.0018152) (0.0000000,-0.0056322) (0.0000000,-0.0136289) (0.0000000,-0.0221642) (0.0000000,-0.0312277)
    (0.0000000,-0.0408100) (0.0000000,-0.0509022) (0.0000000,-0.0614959) (0.0000000,-0.0725832) (0.0000000,-0.0841567)
    (0.0000000,-0.0962094) (0.0000000,-0.1087345) (0.0000000,-0.1217259) (0.0000000,-0.1351775) (0.0000000,-0.1490837)
    (0.0000000,-0.1634389) (0.0000000,-0.1782382) (0.0000000,-0.1934765) (0.0000000,-0.2091492) (0.0000000,-0.2252518)
    (0.0000000,-0.2417800)
  };
  \draw[gray!70,thin] plot coordinates {
    (-0.0000000,0.0000000) (-0.0249499,0.0091187) (-0.0497995,0.0157187) (-0.0745482,0.0206620) (-0.0991957,0.0242560)
    (-0.1237415,0.0266781) (-0.1481853,0.0280483) (-0.1725266,0.0284542) (-0.1967650,0.0279640) (-0.2209000,0.0266324)
    (-0.2449312,0.0245044) (-0.2688582,0.0216184) (-0.2926804,0.0180070) (-0.3163975,0.0136988) (-0.3400089,0.0087190)
    (-0.3635142,0.0030899) (-0.3869129,-0.0031683) (-0.4102045,-0.0100377) (-0.4333886,-0.0175017) (-0.4564645,-0.0255453)
    (-0.4794318,-0.0341546) (-0.5022901,-0.0433169) (-0.5250387,-0.0530204) (-0.5476771,-0.0632541) (-0.5702049,-0.0740078)
    (-0.5926213,-0.0852719) (-0.6149260,-0.0970375) (-0.6371184,-0.1092962) (-0.6591977,-0.1220400) (-0.6811636,-0.1352615)
    (-0.7030153,-0.1489536) (-0.7247524,-0.1631096) (-0.7463741,-0.1777232) (-0.7678799,-0.1927883) (-0.7892691,-0.2082992)
    (-0.8105411,-0.2242504) (-0.8316953,-0.2406368) (-0.8527310,-0.2574533) (-0.8736474,-0.2746953) (-0.8944440,-0.2923582)
    (-0.9151201,-0.3104376)
  };
  \draw[gray!70,thin] plot coordinates {
    (-0.0000000,0.0000000) (-0.0432146,0.0040046) (-0.0862552,0.0055111) (-0.1291212,0.0053816) (-0.1718120,0.0039234)
    (-0.2143266,0.0013143) (-0.2566645,-0.0023259) (-0.2988249,-0.0069093) (-0.3408070,-0.0123677) (-0.3826101,-0.0186464)
    (-0.4242334,-0.0257002) (-0.4656761,-0.0334906) (-0.5069374,-0.0419849) (-0.5480166,-0.0511545) (-0.5889127,-0.0609740)
    (-0.6296251,-0.0714211) (-0.6701528,-0.0824755) (-0.7104951,-0.0941191) (-0.7506510,-0.1063352) (-0.7906197,-0.1191087)
    (-0.8304003,-0.1324258) (-0.8699919,-0.1462734) (-0.9093937,-0.1606398) (-0.9486046,-0.1755138) (-0.9876238,-0.1908851)
    (-1.0264503,-0.2067440) (-1.0650831,-0.2230815) (-1.1035214,-0.2398891) (-1.1417640,-0.2571586) (-1.1798099,-0.2748825)
    (-1.2176583,-0.2930537) (-1.2553079,-0.3116652) (-1.2927579,-0.3307106) (-1.3300070,-0.3501839) (-1.3670542,-0.3700790)
    (-1.4038984,-0.3903905) (-1.4405385,-0.4111129) (-1.4769733,-0.4322412) (-1.5132017,-0.4537705) (-1.5492225,-0.4756962)
    (-1.5850345,-0.4980136)
  };
  \draw[gray!70,thin] plot coordinates {
    (-0.0000000,0.0000000) (-0.0498999,-0.0029814) (-0.0995989,-0.0084327) (-0.1490964,-0.0154919) (-0.1983914,-0.0238514)
    (-0.2474831,-0.0333333) (-0.2963707,-0.0438178) (-0.3450533,-0.0552167) (-0.3935300,-0.0674619) (-0.4418001,-0.0804984)
    (-0.4898625,-0.0942809) (-0.5377164,-0.1087709) (-0.5853609,-0.1239355) (-0.6327950,-0.1397458) (-0.6800178,-0.1561765)
    (-0.7270284,-0.1732051) (-0.7738258,-0.1908111) (-0.8204091,-0.2089763) (-0.8667771,-0.2276840) (-0.9129290,-0.2469188)
    (-0.9588637,-0.2666667) (-1.0045802,-0.2869146) (-1.0500774,-0.3076506) (-1.0953542,-0.3288634) (-1.1404097,-0.3505424)
    (-1.1852427,-0.3726780) (-1.2298521,-0.3952608) (-1.2742367,-0.4182822) (-1.3183955,-0.4417340) (-1.3623272,-0.4656083)
    (-1.4060307,-0.4898979) (-1.4495047,-0.5145959) (-1.4927482,-0.5396954) (-1.5357598,-0.5651902) (-1.5785382,-0.5910744)
    (-1.6210823,-0.6173420) (-1.6633906,-0.6439876) (-1.7054619,-0.6710059) (-1.7472948,-0.6983918) (-1.7888881,-0.7261405)
    (-1.8302401,-0.7542472)
  };
  \draw[gray!70,thin] plot coordinates {
    (-0.0000000,-0.0000000) (-0.0432146,-0.0099674) (-0.0862552,-0.0223766) (-0.1291212,-0.0363654) (-0.1718120,-0.0516262)
    (-0.2143266,-0.0679810) (-0.2566645,-0.0853097) (-0.2988249,-0.1035242) (-0.3408070,-0.1225561) (-0.3826101,-0.1423505)
    (-0.4242334,-0.1628617) (-0.4656761,-0.1840512) (-0.5069374,-0.2058860) (-0.5480166,-0.2283371) (-0.5889127,-0.2513790)
    (-0.6296251,-0.2749891) (-0.6701528,-0.2991468) (-0.7104951,-0.3238336) (-0.7506510,-0.3490328) (-0.7906197,-0.3747288)
    (-0.8304003,-0.4009076) (-0.8699919,-0.4275558) (-0.9093937,-0.4546614) (-0.9486046,-0.4822130) (-0.9876238,-0.5101998)
    (-1.0264503,-0.5386120) (-1.0650831,-0.5674401) (-1.1035214,-0.5966753) (-1.1417640,-0.6263093) (-1.1798099,-0.6563341)
    (-1.2176583,-0.6867422) (-1.2553079,-0.7175265) (-1.2927579,-0.7486801) (-1.3300070,-0.7801966) (-1.3670542,-0.8120697)
    (-1.4038984,-0.8442935) (-1.4405385,-0.8768623) (-1.4769733,-0.9097705) (-1.5132017,-0.9430131) (-1.5492225,-0.9765848)
    (-1.5850345,-1.0104809)
  };
  \draw[gray!70,thin] plot coordinates {
    (-0.0000000,-0.0000000) (-0.0249499,-0.0150815) (-0.0497995,-0.0325842) (-0.0745482,-0.0516459) (-0.0991957,-0.0719587)
    (-0.1237415,-0.0933448) (-0.1481853,-0.1156839) (-0.1725266,-0.1388877) (-0.1967650,-0.1628879) (-0.2209000,-0.1876293)
    (-0.2449312,-0.2130662) (-0.2688582,-0.2391602) (-0.2926804,-0.2658779) (-0.3163975,-0.2931904) (-0.3400089,-0.3210721)
    (-0.3635142,-0.3495001) (-0.3869129,-0.3784539) (-0.4102045,-0.4079150) (-0.4333886,-0.4378663) (-0.4564645,-0.4682923)
    (-0.4794318,-0.4991788) (-0.5022901,-0.5305124) (-0.5250387,-0.5622808) (-0.5476771,-0.5944727) (-0.5702049,-0.6270771)
    (-0.5926213,-0.6600841) (-0.6149260,-0.6934841) (-0.6371184,-0.7272682) (-0.6591977,-0.7614279) (-0.6811636,-0.7959551)
    (-0.7030153,-0.8308423) (-0.7247524,-0.8660821) (-0.7463741,-0.9016676) (-0.7678799,-0.9375922) (-0.7892691,-0.9738495)
    (-0.8105411,-1.0104335) (-0.8316953,-1.0473384) (-0.8527310,-1.0845584) (-0.8736474,-1.1220883) (-0.8944440,-1.1599228)
    (-0.9151201,-1.1980569)
  };
  \draw[gray!70,thin] plot coordinates {
    (-0.0000000,-0.0000000) (-0.0000000,-0.0169534) (-0.0000000,-0.0363204) (-0.0000000,-0.0572389) (-0.0000000,-0.0794010)
    (-0.0000000,-0.1026286) (-0.0000000,-0.1268016) (-0.0000000,-0.1518317) (-0.0000000,-0.1776503) (-0.0000000,-0.2042025)
    (-0.0000000,-0.2314424) (-0.0000000,-0.2593315) (-0.0000000,-0.2878365) (-0.0000000,-0.3169284) (-0.0000000,-0.3465815)
    (-0.0000000,-0.3767730) (-0.0000000,-0.4074824) (-0.0000000,-0.4386909) (-0.0000000,-0.4703816) (-0.0000000,-0.5025389)
    (-0.0000000,-0.5351485) (-0.0000000,-0.5681971) (-0.0000000,-0.6016723) (-0.0000000,-0.6355626) (-0.0000000,-0.6698572)
    (-0.0000000,-0.7045460) (-0.0000000,-0.7396194) (-0.0000000,-0.7750685) (-0.0000000,-0.8108847) (-0.0000000,-0.8470599)
    (-0.0000000,-0.8835865) (-0.0000000,-0.9204572) (-0.0000000,-0.9576649) (-0.0000000,-0.9952030) (-0.0000000,-1.0330651)
    (-0.0000000,-1.0712450) (-0.0000000,-1.1097369) (-0.0000000,-1.1485352) (-0.0000000,-1.1876344) (-0.0000000,-1.2270291)
    (-0.0000000,-1.2667145)
  };
  \draw[gray!70,thin] plot coordinates {
    (0.0000000,-0.0000000) (0.0249499,-0.0150815) (0.0497995,-0.0325842) (0.0745482,-0.0516459) (0.0991957,-0.0719587)
    (0.1237415,-0.0933448) (0.1481853,-0.1156839) (0.1725266,-0.1388877) (0.1967650,-0.1628879) (0.2209000,-0.1876293)
    (0.2449312,-0.2130662) (0.2688582,-0.2391602) (0.2926804,-0.2658779) (0.3163975,-0.2931904) (0.3400089,-0.3210721)
    (0.3635142,-0.3495001) (0.3869129,-0.3784539) (0.4102045,-0.4079150) (0.4333886,-0.4378663) (0.4564645,-0.4682923)
    (0.4794318,-0.4991788) (0.5022901,-0.5305124) (0.5250387,-0.5622808) (0.5476771,-0.5944727) (0.5702049,-0.6270771)
    (0.5926213,-0.6600841) (0.6149260,-0.6934841) (0.6371184,-0.7272682) (0.6591977,-0.7614279) (0.6811636,-0.7959551)
    (0.7030153,-0.8308423) (0.7247524,-0.8660821) (0.7463741,-0.9016676) (0.7678799,-0.9375922) (0.7892691,-0.9738495)
    (0.8105411,-1.0104335) (0.8316953,-1.0473384) (0.8527310,-1.0845584) (0.8736474,-1.1220883) (0.8944440,-1.1599228)
    (0.9151201,-1.1980569)
  };
  \draw[gray!70,thin] plot coordinates {
    (0.0000000,-0.0000000) (0.0432146,-0.0099674) (0.0862552,-0.0223766) (0.1291212,-0.0363654) (0.1718120,-0.0516262)
    (0.2143266,-0.0679810) (0.2566645,-0.0853097) (0.2988249,-0.1035242) (0.3408070,-0.1225561) (0.3826101,-0.1423505)
    (0.4242334,-0.1628617) (0.4656761,-0.1840512) (0.5069374,-0.2058860) (0.5480166,-0.2283371) (0.5889127,-0.2513790)
    (0.6296251,-0.2749891) (0.6701528,-0.2991468) (0.7104951,-0.3238336) (0.7506510,-0.3490328) (0.7906197,-0.3747288)
    (0.8304003,-0.4009076) (0.8699919,-0.4275558) (0.9093937,-0.4546614) (0.9486046,-0.4822130) (0.9876238,-0.5101998)
    (1.0264503,-0.5386120) (1.0650831,-0.5674401) (1.1035214,-0.5966753) (1.1417640,-0.6263093) (1.1798099,-0.6563341)
    (1.2176583,-0.6867422) (1.2553079,-0.7175265) (1.2927579,-0.7486801) (1.3300070,-0.7801966) (1.3670542,-0.8120697)
    (1.4038984,-0.8442935) (1.4405385,-0.8768623) (1.4769733,-0.9097705) (1.5132017,-0.9430131) (1.5492225,-0.9765848)
    (1.5850345,-1.0104809)
  };
  \draw[thick] plot coordinates {
    (0.0000000,-0.0000000) (0.0498999,-0.0029814) (0.0995989,-0.0084327) (0.1490964,-0.0154919) (0.1983914,-0.0238514)
    (0.2474831,-0.0333333) (0.2963707,-0.0438178) (0.3450533,-0.0552167) (0.3935300,-0.0674619) (0.4418001,-0.0804984)
    (0.4898625,-0.0942809) (0.5377164,-0.1087709) (0.5853609,-0.1239355) (0.6327950,-0.1397458) (0.6800178,-0.1561765)
    (0.7270284,-0.1732051) (0.7738258,-0.1908111) (0.8204091,-0.2089763) (0.8667771,-0.2276840) (0.9129290,-0.2469188)
    (0.9588637,-0.2666667) (1.0045802,-0.2869146) (1.0500774,-0.3076506) (1.0953542,-0.3288634) (1.1404097,-0.3505424)
    (1.1852427,-0.3726780) (1.2298521,-0.3952608) (1.2742367,-0.4182822) (1.3183955,-0.4417340) (1.3623272,-0.4656083)
    (1.4060307,-0.4898979) (1.4495047,-0.5145959) (1.4927482,-0.5396954) (1.5357598,-0.5651902) (1.5785382,-0.5910744)
    (1.6210823,-0.6173420) (1.6633906,-0.6439876) (1.7054619,-0.6710059) (1.7472948,-0.6983918) (1.7888881,-0.7261405)
    (1.8302401,-0.7542472)
  };
  \draw[fill=white,thick] (0,0.25) circle (0.25);
  \draw[gray] (-0.25,0.25) arc[start angle=180,end angle=360,x radius=0.25,y radius=0.065];
  \fill (0,0.25) circle (1pt);
  \draw (0.07,0.32) -- (0.62,0.65) node[right] {$G(0)$};
  \fill (0,0) circle (1pt) node[below left] {$P_0$};
  \node at (0,-1.50) {\small (a) Perspective view};
\end{tikzpicture}
\end{minipage}\hfill
\begin{minipage}[b]{0.50\textwidth}
\centering
\begin{tikzpicture}[scale=0.84,>=Latex]
  \path[use as bounding box] (-3.3,-2.4) rectangle (3.5,1.8);
  \draw[->] (-3.1,0) -- (3.1,0) node[right] {$x$};
  \draw[->] (0,-1.85) -- (0,1.5) node[above] {$z$};
  \draw[thick] plot coordinates {
    (-2.5000000,-1.6601958) (-2.4579832,-1.6185184) (-2.4159664,-1.5771958) (-2.3739496,-1.5362310) (-2.3319328,-1.4956271)
    (-2.2899160,-1.4553874) (-2.2478992,-1.4155152) (-2.2058824,-1.3760139) (-2.1638655,-1.3368870) (-2.1218487,-1.2981382)
    (-2.0798319,-1.2597711) (-2.0378151,-1.2217897) (-1.9957983,-1.1841979) (-1.9537815,-1.1469996) (-1.9117647,-1.1101993)
    (-1.8697479,-1.0738012) (-1.8277311,-1.0378097) (-1.7857143,-1.0022297) (-1.7436975,-0.9670657) (-1.7016807,-0.9323230)
    (-1.6596639,-0.8980065) (-1.6176471,-0.8641218) (-1.5756303,-0.8306743) (-1.5336134,-0.7976698) (-1.4915966,-0.7651144)
    (-1.4495798,-0.7330144) (-1.4075630,-0.7013763) (-1.3655462,-0.6702069) (-1.3235294,-0.6395135) (-1.2815126,-0.6093034)
    (-1.2394958,-0.5795847) (-1.1974790,-0.5503654) (-1.1554622,-0.5216544) (-1.1134454,-0.4934608) (-1.0714286,-0.4657943)
    (-1.0294118,-0.4386650) (-0.9873950,-0.4120839) (-0.9453782,-0.3860625) (-0.9033613,-0.3606131) (-0.8613445,-0.3357489)
    (-0.8193277,-0.3114840) (-0.7773109,-0.2878336) (-0.7352941,-0.2648140) (-0.6932773,-0.2424431) (-0.6512605,-0.2207402)
    (-0.6092437,-0.1997266) (-0.5672269,-0.1794256) (-0.5252101,-0.1598633) (-0.4831933,-0.1410687) (-0.4411765,-0.1230744)
    (-0.3991597,-0.1059179) (-0.3571429,-0.0896421) (-0.3151261,-0.0742978) (-0.2731092,-0.0599451) (-0.2310924,-0.0466582)
    (-0.1890756,-0.0345305) (-0.1470588,-0.0236857) (-0.1050420,-0.0142986) (-0.0630252,-0.0066454) (-0.0210084,-0.0012789)
    (0.0210084,-0.0012789) (0.0630252,-0.0066454) (0.1050420,-0.0142986) (0.1470588,-0.0236857) (0.1890756,-0.0345305)
    (0.2310924,-0.0466582) (0.2731092,-0.0599451) (0.3151261,-0.0742978) (0.3571429,-0.0896421) (0.3991597,-0.1059179)
    (0.4411765,-0.1230744) (0.4831933,-0.1410687) (0.5252101,-0.1598633) (0.5672269,-0.1794256) (0.6092437,-0.1997266)
    (0.6512605,-0.2207402) (0.6932773,-0.2424431) (0.7352941,-0.2648140) (0.7773109,-0.2878336) (0.8193277,-0.3114840)
    (0.8613445,-0.3357489) (0.9033613,-0.3606131) (0.9453782,-0.3860625) (0.9873950,-0.4120839) (1.0294118,-0.4386650)
    (1.0714286,-0.4657943) (1.1134454,-0.4934608) (1.1554622,-0.5216544) (1.1974790,-0.5503654) (1.2394958,-0.5795847)
    (1.2815126,-0.6093034) (1.3235294,-0.6395135) (1.3655462,-0.6702069) (1.4075630,-0.7013763) (1.4495798,-0.7330144)
    (1.4915966,-0.7651144) (1.5336134,-0.7976698) (1.5756303,-0.8306743) (1.6176471,-0.8641218) (1.6596639,-0.8980065)
    (1.7016807,-0.9323230) (1.7436975,-0.9670657) (1.7857143,-1.0022297) (1.8277311,-1.0378097) (1.8697479,-1.0738012)
    (1.9117647,-1.1101993) (1.9537815,-1.1469996) (1.9957983,-1.1841979) (2.0378151,-1.2217897) (2.0798319,-1.2597711)
    (2.1218487,-1.2981382) (2.1638655,-1.3368870) (2.2058824,-1.3760139) (2.2478992,-1.4155152) (2.2899160,-1.4553874)
    (2.3319328,-1.4956271) (2.3739496,-1.5362310) (2.4159664,-1.5771958) (2.4579832,-1.6185184) (2.5000000,-1.6601958)
  };
  \fill (0,0) circle (1.5pt) node[below right] {$P_0$};
  \draw[thick,fill=white] (0,0.46) circle (0.46);
  \fill (0,0.46) circle (1.4pt);
  \draw (0.08,0.52) -- (0.72,1.0) node[right] {$G(0)$};
  \draw[<->] (-0.62,0) -- (-0.62,0.46) node[midway,left] {$\rho$};
  \draw[->,dashed] (1.6,-0.72) -- (2.2,-1.26) node[midway,right] {$r$};
  \node at (0,-2.2) {\small (b) Meridional section};
\end{tikzpicture}
\end{minipage}
\caption{A finite sphere at rest at the apex of the standard Norton dome:
(a) a perspective view of the surface of revolution and (b) a schematic
meridional section. Here $r$ is arclength along the support, not the horizontal
distance $x$. The sphere center follows an offset curve; its height loss
differs from that of a point particle on the support. Both panels show the
initial configuration, not a sequence of spontaneous departures.}
\label{fig:standard-dome}
\end{figure}

\section{The standard dome: point particle versus finite sphere}
\label{sec:point}
\label{sec:original-proof}

Let $r\ge0$ denote arclength along a meridian of the dome, measured from the
apex, and let $h(r)$ be the vertical drop from the apex.  We write the standard
Norton \cite{Norton2008} (see Fig \ref{fig:standard-dome}) as
\begin{equation}
 h(r)=\frac{2c}{3g}r^{3/2},
 \label{eq:standard-dome}
\end{equation}

where $g>0$ is the gravitational acceleration and the constant $c>0$ restores
the dimensions explicitly, with $[c]=L^{1/2}T^{-2}$. We use only the local
profile $0\le r<(g/c)^2$, where the arclength parametrization is real.
If $\theta(r)$ is the angle between the tangent
and the horizontal, then the equation of motion:
\begin{equation}
mg \sin\theta=mg\frac{dh}{dr}=mg\frac{c}{g}r^{1/2}=ma.
\end{equation}
For a frictionless point particle,
\begin{equation}
 \ddot r=g\sin\theta=c\sqrt r.
 \label{eq:standard-ode}
\end{equation}
The initial conditions are
\begin{equation}
 r(0)=0,
 \qquad
 \dot r(0)=0.
 \label{eq:initial-data}
\end{equation}
Besides the stationary solution $r(t)\equiv0$, Eq.~\eqref{eq:standard-ode}
admits, for every $T\ge0$,
\begin{equation}
 r_T(t)=
 \begin{cases}
 0, & t\le T,\\[1mm]
 \dfrac{c^2}{144}(t-T)^4, & t>T.
 \end{cases}
 \label{eq:standard-delayed}
\end{equation}
Thus the same initial position and velocity admit infinitely many forward
solutions.

Now let us consider the system with finite size. The moving object is now a rigid sphere of radius $\rho>0$ with a
spherically symmetric mass distribution. The case $\rho=0$ denotes the
point-particle model for comparison.  The surface is frictionless and the sphere starts
without spin.  Since both gravity and the normal reaction act through the
sphere center, no spin is generated in this idealized model.

The argument has three steps. We first derive the relation between
the contact arclength $r$ and the center arclength $q$. We then use
Newton's second law to obtain the center speed. Combining these
relations gives the travel-time integrand $dt/dr$, whose behavior
near the apex determines whether departure can occur in finite time.

\paragraph{Step 1: Geometry.}
Let $r$ denote arclength along a meridian of the dome, measured from
the apex (as the origin $(0,0)$), and let $\theta(r)$ be the downward inclination of its tangent.
The contact point has coordinates
\begin{equation}
 \mathbf P(r)=(x(r),-h(r)),
 \qquad
 x'(r)=\cos\theta(r),
 \qquad
 h'(r)=\sin\theta(r).
\end{equation}
For the standard Norton dome,
\begin{equation}
 h(r)=\frac{2c}{3g}r^{3/2},
 \qquad
 \sin\theta(r)=\frac{c}{g}\sqrt r.
\end{equation}
Differentiating the latter identity gives
\begin{equation}
 \cos\theta(r)\frac{d\theta}{dr}
 =\frac{c}{2g\sqrt r}.
\end{equation}
Hence, for $0<r<(g/c)^2$, the meridional curvature is
\begin{equation}
 \kappa(r)=\frac{d\theta}{dr}
 =\frac{c}{2g\sqrt r}
 \left(1-\frac{c^2r}{g^2}\right)^{-1/2}.
 \label{eq:standard-curvature}
\end{equation}

The unit normal pointing from the dome toward the sphere center is
$\mathbf n(r)=(\sin\theta(r),\cos\theta(r))$.
For a sphere of radius $\rho$, the center therefore lies at
\begin{equation}
 \mathbf G(r)=\mathbf P(r)+\rho\mathbf n(r).
\end{equation}
Differentiating with respect to $r$ yields
\begin{align}
 \frac{d\mathbf G}{dr}
 &=(\cos\theta,-\sin\theta)
   +\rho\frac{d\theta}{dr}(\cos\theta,-\sin\theta)\\
 &=(1+\rho\kappa)(\cos\theta,-\sin\theta).
\end{align}
Let $q$ denote arclength along the center trajectory, with $q(0)=0$.
Since $\kappa(r)>0$,
\begin{equation}
 \frac{dq}{dr}
 =\left|\frac{d\mathbf G}{dr}\right|
 =1+\rho\kappa(r)
 =1+\frac{c\rho}{2g\sqrt r}
 \left(1-\frac{c^2r}{g^2}\right)^{-1/2}.
 \label{eq:standard-qprime}
\end{equation}
The unit tangent to the center trajectory is consequently
\begin{equation}
 \mathbf T=\frac{d\mathbf G}{dq}
 =(\cos\theta,-\sin\theta).
\end{equation}

The vertical coordinate of the center is
$G_z(r)=-h(r)+\rho\cos\theta(r)$.
Since $G_z(0)=\rho$, its height loss from the initial position is $ \Delta h_\rho(r)=\rho-G_z(r)$
\begin{equation}
 \Delta h_\rho(r)
 =h(r)+\rho(1-\cos\theta(r))
 =\frac{2c}{3g}r^{3/2}
 +\rho\left(1-\sqrt{1-\frac{c^2r}{g^2}}\right).
 \label{eq:standard-height}
\end{equation}
\paragraph{Step 2: Newton's second law and energy.}
For frictionless contact, the normal reaction has no component along
the tangent $\mathbf T$. Newton's second law therefore gives
\begin{equation}
 m\ddot q=mg\sin\theta(r).
\end{equation}
We now integrate this equation to express the center speed in terms
of the center's height loss.

Multiplying by $\dot q$ and dividing by $m$ yields
\begin{equation}
 \frac{d}{dt}\left(\frac12\dot q^{\,2}\right)
 =g\sin\theta\,\dot q.
\end{equation}
To rewrite the right-hand side, differentiate the center-height loss
$\Delta h_\rho=h(r)+\rho(1-\cos\theta(r))$ with respect to time:
\begin{align}
 \frac{d\Delta h_\rho}{dt}
 &=h'(r)\dot r
   +\rho\sin\theta\,\theta'(r)\dot r\\
 &=\sin\theta(1+\rho\kappa)\dot r\\
 &=\sin\theta\,\dot q,
\end{align}
where we used $h'=\sin\theta$, $\theta'=\kappa$, and
$\dot q=(1+\rho\kappa)\dot r$. Consequently,
\begin{equation}
 \frac{d}{dt}
 \left(\frac12\dot q^{\,2}-g\Delta h_\rho\right)=0.
\end{equation}
The sphere starts from rest at the apex, where both $\dot q$ and
$\Delta h_\rho$ vanish. Hence
\begin{equation}
 \frac12m\dot q^{\,2}=mg\Delta h_\rho(r).
\end{equation}
This is the conversion of gravitational potential energy into
translational kinetic energy; the sphere does not acquire spin
under the frictionless assumptions above. On any hypothetical
outward branch, $\dot q>0$, so
\begin{equation}
 \dot q=\sqrt{2g\Delta h_\rho(r)}.
\end{equation}

\paragraph{Step 3: The travel-time integrand $\frac{dt}{dr}$.}
The chain rule gives
\begin{equation}
 \dot q=\frac{dq}{dr}\dot r
       =(1+\rho\kappa(r))\dot r.
\end{equation}
Consequently, on an outward branch with $r>0$, using Eq.~\eqref{eq:standard-qprime} and \eqref{eq:standard-curvature}.
\begin{equation}
 \frac{dt}{dr}
 = \frac{\frac{dt}{dq}}{\frac{dr}{dq}}=\frac{\frac{dq}{dr}}{\dot q}=\frac{1+\rho\kappa(r)}{\sqrt{2g\Delta h_\rho(r)}}
 =:I_\rho(r).
 \label{eq:standard-integrand-def}
\end{equation}
Substituting Eqs.~\eqref{eq:standard-curvature} and
\eqref{eq:standard-height} yields
\begin{equation}
 \frac{dt}{dr}=I_\rho(r)=
 \frac{
 1+\dfrac{c\rho}{2g\sqrt r}
 \left(1-\dfrac{c^2r}{g^2}\right)^{-1/2}
 }{
 \sqrt{2g\left[
 \dfrac{2c}{3g}r^{3/2}
 +\rho\left(1-\sqrt{1-\dfrac{c^2r}{g^2}}\right)
 \right]}
 }.
 \label{eq:standard-integrand}
\end{equation}
The following theorem shows that, for every fixed $\rho>0$,
the integral of $I_\rho(r)$ diverges at the apex. Thus no outward
branch can connect initial rest at the apex to a positive distance
in finite time.

Figure~\ref{fig:standard-dome} illustrates the initial sphere geometry used in Theorem~\ref{thm:standard-piecewise}.

\begin{theorem}[Point particle versus finite sphere on the standard Norton dome]
\label{thm:standard-piecewise}
Fix a sufficiently small $\varepsilon>0$.  The travel time from the exact apex
to $r=\varepsilon$ is
\begin{equation}
 \Tcal_\rho(\varepsilon)
 :=\lim_{\delta\to0^+}
 \int_\delta^\varepsilon I_\rho(r)\,dr.
\end{equation}
It obeys the piecewise classification
\begin{equation}
 \Tcal_\rho(\varepsilon)=
 \begin{cases}
 +\infty, & \rho>0,\\[2mm]
 2\sqrt{\dfrac{3}{c}}\,\varepsilon^{1/4}, & \rho=0.
 \end{cases}
 \label{eq:standard-travel-piecewise}
\end{equation}
Consequently, every fixed positive sphere radius excludes finite-time
spontaneous departure from exact rest, whereas the point-particle model admits
the delayed solutions \eqref{eq:standard-delayed}.
\end{theorem}

\begin{proof}
For fixed $\rho>0$, Eq.~\eqref{eq:standard-integrand} gives
\begin{equation}
 \lim_{r\to0^+}rI_\rho(r)
 =\frac12\sqrt{\frac{\rho}{g}}>0.
 \label{eq:standard-log-asymptotic}
\end{equation}
Thus $I_\rho(r)$ behaves as a positive constant times $1/r$
near the apex, and
\begin{equation}
 \int_0^\varepsilon I_\rho(r)\,dr=+\infty.
\end{equation}
Hence a sphere of any fixed positive radius cannot depart
from exact rest at the apex in finite time.

For $\rho=0$, Eq.~\eqref{eq:standard-integrand} reduces exactly to
\begin{equation}
 I_0(r)=\frac{\sqrt3}{2\sqrt c}\,r^{-3/4}.
\end{equation}
Hence
\begin{equation}
 \Tcal_0(\varepsilon)
 =\frac{\sqrt3}{2\sqrt c}
 \int_0^\varepsilon r^{-3/4}\,dr
 =2\sqrt{\frac3c}\,\varepsilon^{1/4}<\infty.
\end{equation}
Inverting this relation after an arbitrary waiting time $T$ yields
Eq.~\eqref{eq:standard-delayed}.
\end{proof}

When $\rho>0$, the key point is not rotational inertia. The positive radius changes both the
center trajectory and its vertical drop. Near the apex the numerator of the
travel-time integrand acquires an $r^{-1/2}$ factor, while the center speed is
of order $r^{1/2}$; their ratio is therefore of order $1/r$. This construction also allows pure rolling. If a sphere rolls without slipping and has moment of inertia
$I=\eta m\rho^2$, then
\begin{equation}
 \frac12m(1+\eta)\dot q^{\,2}=mg\Delta h_\rho.
\end{equation}
The travel-time integrand is merely multiplied by $\sqrt{1+\eta}$.  The
logarithmic divergence is unchanged. Thus rotational inertia does not change the conclusion; the finite contact geometry already does so.

\section{Generalized Norton family}
\label{sec:generalized}

We now extend the sphere calculation to a family of dome profiles
whose point-particle dynamics follow a power law \cite{Berry2025}.
Consider the one-parameter family
\begin{equation}
 h_a(r)=\frac{a}{a+1}\frac{c}{g}
 r^{(a+1)/a},
 \qquad 0<a<\infty.
 \label{eq:general-profile}
\end{equation}
Here $[c]=L^{1-1/a}T^{-2}$, so its dimensions depend on $a$.
The geometric model is restricted to $0\le r<R_a$, where $R_a=(g/c)^a$;
all departure statements concern a sufficiently small neighborhood of the apex.
The apex geometry changes at $a=1$: the curvature diverges for $a>1$,
is finite and nonzero for $a=1$, and vanishes for $0<a<1$.  The dynamical
classification below is governed by the same threshold. The tangent angle
satisfies
\begin{equation}
 \sin\theta(r)=\frac{dh_a}{dr}
 =\frac{c}{g}r^{1/a},
\end{equation}
so the point-particle equation is
\begin{equation}
 \ddot r=cr^{1/a}.
 \label{eq:general-point-ode}
\end{equation}

\subsection{Point-particle classification}
For every finite $a>0$, consider the point-particle equation
\eqref{eq:general-point-ode} with $r(0)=\dot r(0)=0$ and $r\ge0$.
The forward radial solutions are classified as follows:
\begin{equation}
 \Scal_{a,0}=
 \begin{cases}
 \{r(t)\equiv0\}, & 0<a\le1,\\[2mm]
 \{r(t)\equiv0\}\cup\{r_{T,a}:T\ge0\}, & a>1,
 \end{cases}
 \label{eq:point-classification}
\end{equation}
where, for $a>1$,
\begin{equation}
 r_{T,a}(t)=
 \begin{cases}
 0,&t\le T,\\[2mm]
 \displaystyle
 \left[
 \frac{c(a-1)^2}{2a(a+1)}
 \right]^{a/(a-1)}
 (t-T)^{2a/(a-1)},&t>T.
 \end{cases}
 \label{eq:general-delayed}
\end{equation}

Since $\ddot r\ge0$ and $\dot r(0)=0$, every solution is nondecreasing.
Its zero set is an initial interval; once $r>0$, its velocity is positive.
Thus every nonstationary solution consists of a waiting interval followed by
an outward branch. Multiplying Eq.~\eqref{eq:general-point-ode} by $\dot r$ and using the zero
initial energy gives, on an outward-moving branch,
\begin{equation}
 \frac12\dot r^{\,2}
 =\frac{ac}{a+1}r^{(a+1)/a}.
\end{equation}
Hence
\begin{equation}
 \frac{dt}{dr}
 =\sqrt{\frac{a+1}{2ac}}
 r^{-(a+1)/(2a)}.
 \label{eq:general-point-travel}
\end{equation}
The integral from $0$ to $\varepsilon$ is finite exactly when
\begin{equation}
 \frac{a+1}{2a}<1,
\end{equation}
which is equivalent to $a>1$.  For $0<a\le1$, finite-time departure is
therefore impossible.  For $a>1$, integration and inversion give
Eq.~\eqref{eq:general-delayed}.  Since $2a/(a-1)>2$, the departing branch joins
the stationary branch with matching position, velocity, and acceleration.

\subsection{Finite sphere for arbitrary finite \texorpdfstring{$a>0$}{a>0}}

Applying the preceding construction to the generalized profile gives
\begin{equation}
 \kappa_a(r)
 =\frac{c}{ag}
 \frac{r^{1/a-1}}
 {\sqrt{1-\dfrac{c^2}{g^2}r^{2/a}}},
 \label{eq:general-curvature}
\end{equation}
and
\begin{equation}
 \frac{dq}{dr}=1+\rho\kappa_a(r).
 \label{eq:general-qprime}
\end{equation}
Integrating with $q(0)=0$ yields
\begin{equation}
 q(r)=r+\rho\arcsin\!\left(\frac{c}{g}r^{1/a}\right).
 \label{eq:qexact}
\end{equation}
The center-height loss is
\begin{equation}
 \Delta h_{a,\rho}(r)
 =\frac{a}{a+1}\frac{c}{g}r^{(a+1)/a}
 +\rho\left(1-\sqrt{1-\frac{c^2}{g^2}r^{2/a}}\right),
 \label{eq:general-sphere-height}
\end{equation}
so the travel-time integrand is
\begin{equation}
 I_{a,\rho}(r)
 :=\frac{dt}{dr}
 =\frac{1+\rho\kappa_a(r)}
 {\sqrt{2g\Delta h_{a,\rho}(r)}}.
 \label{eq:general-sphere-integrand}
\end{equation}

\begin{theorem}[Finite-radius classification]
\label{thm:general-sphere}
For every fixed $\rho>0$ and finite $a>0$,
\begin{equation}
 \int_0^\varepsilon I_{a,\rho}(r)\,dr=+\infty
\end{equation}
for every sufficiently small $\varepsilon>0$. Hence the sphere
cannot spontaneously depart from exact rest at the apex.
\end{theorem}

\begin{proof}
Taking the leading terms in
Eq.~\eqref{eq:general-sphere-integrand} as $r\to0^+$ gives
\begin{equation}
 I_{a,\rho}(r)\sim
 \begin{cases}
 \dfrac1a\sqrt{\dfrac{\rho}{g}}\,r^{-1},
 & a>1,\\[3mm]
 \sqrt{\dfrac1c+\dfrac{\rho}{g}}\,r^{-1},
 & a=1,\\[3mm]
 \sqrt{\dfrac{a+1}{2ac}}\,r^{-(a+1)/(2a)},
 & 0<a<1.
 \end{cases}
 \label{eq:general-asymptotics}
\end{equation}
The first two cases diverge logarithmically upon integration.
In the third case, $(a+1)/(2a)>1$, so the integral also diverges.
\end{proof}

Together with Eq.~\eqref{eq:point-classification}, this gives
\begin{equation}
 \Scal_{a,\rho}=
 \begin{cases}
 \{0\}, & \rho>0\ \text{or}\ 0<a\le1,\\[1mm]
 \{0\}\cup\{r_{T,a}:T\ge0\},
 & \rho=0\ \text{and}\ a>1,
 \end{cases}
 \label{eq:full-classification}
\end{equation}
within the frictionless radial model.

\paragraph{Universal quadratic center geometry.}
For $a>1$, Eqs.~\eqref{eq:qexact} and
\eqref{eq:general-sphere-height} give
\begin{equation}
 q(r)\sim\frac{\rho c}{g}r^{1/a},
 \qquad
 \Delta h_{a,\rho}(r)\sim
 \frac{\rho c^2}{2g^2}r^{2/a}.
\end{equation}
Eliminating $r$ yields
\begin{equation}
\Delta h_{a,\rho}\sim\frac{q^2}{2\rho},
 \label{eq:universal-quadratic}
\end{equation}
and therefore, on a hypothetical outward branch,
\begin{equation}
 |\dot q|\sim\sqrt{\frac{g}{\rho}}\,q.
 \label{eq:universal-linear-speed}
\end{equation}
Thus the leading center-height law is independent of the exponent
throughout the nonunique point-particle regime.

\paragraph{A direct acceleration bound.}
The tangential equation $\ddot q=f(q)$, with
$f(q)=g\sin\theta(r(q))$, also gives
\begin{equation}
 \frac{df}{dq}
 =\frac{g\cos\theta\,\kappa_a}{1+\rho\kappa_a},
 \qquad
 0\le\frac{df}{dq}\le\frac{g}{\rho}.
 \label{eq:lipschitz-force}
\end{equation}
Since $f(0)=0$, its odd extension across the apex is Lipschitz.
The initial state $(q,\dot q)=(0,0)$ therefore has the unique
solution of rest. For $a>1$, $df/dq\to g/\rho$, giving
$\ddot q\simeq(g/\rho)q$ near the apex: the equilibrium is unique
but unstable.

Contact is locally consistent with this analysis. The center-path
curvature satisfies
$\kappa_q=\kappa_a/(1+\rho\kappa_a)\le1/\rho$, so the normal reaction
on the hypothetical zero-energy outward branch obeys
\begin{equation}
 N=m\left(g\cos\theta
 -2g\Delta h_{a,\rho}\kappa_q\right)\longrightarrow mg
 \qquad(r\to0^+).
\end{equation}
Thus the reaction remains compressive sufficiently close to the apex.

\section{A center-of-mass barrier theorem}
\label{sec:com-barrier}

The sphere calculation suggests a geometric condition independent of shape.
The body need not be spherical and, at this level, need not even be rigid.
The only geometric input is that its center of mass cannot move closer to the
original apex than it was initially. We now formalize it.

Let $P_0$ be a fixed point in space, interpreted as the original apex and
initial contact point.  Choose coordinates with
\begin{equation}
 P_0=0,
 \qquad
 \eZ=(0,0,1),
\end{equation}
and let the uniform gravitational field be $-g\eZ$.  Let $G(t)$ be the center
of mass of a mechanical system of total mass $m>0$, let $K(t)$ denote its total
kinetic energy, and suppose
\begin{equation}
 G(0)=l\eZ,
 \qquad l>0.
 \label{eq:initial-com-distance}
\end{equation}
Define
\begin{equation}
 \mathbf u(t)=G(t)-G(0),
 \qquad
 D(t)=\norm{\mathbf u(t)},
 \qquad
 H(t)=G_z(0)-G_z(t)=-u_z(t).
 \label{eq:com-variables}
\end{equation}

\begin{figure}[H]
\centering
\begin{tikzpicture}[scale=1.05,>=Latex]
  \coordinate (P) at (0,0);
  \draw[thick] (P) circle (1.8);
  \draw[thick] plot coordinates {
    (-2.7000000,-0.8748000) (-2.6454545,-0.8398116) (-2.5909091,-0.8055372) (-2.5363636,-0.7719769) (-2.4818182,-0.7391306)
    (-2.4272727,-0.7069983) (-2.3727273,-0.6755802) (-2.3181818,-0.6448760) (-2.2636364,-0.6148860) (-2.2090909,-0.5856099)
    (-2.1545455,-0.5570479) (-2.1000000,-0.5292000) (-2.0454545,-0.5020661) (-1.9909091,-0.4756463) (-1.9363636,-0.4499405)
    (-1.8818182,-0.4249488) (-1.8272727,-0.4006711) (-1.7727273,-0.3771074) (-1.7181818,-0.3542579) (-1.6636364,-0.3321223)
    (-1.6090909,-0.3107008) (-1.5545455,-0.2899934) (-1.5000000,-0.2700000) (-1.4454545,-0.2507207) (-1.3909091,-0.2321554)
    (-1.3363636,-0.2143041) (-1.2818182,-0.1971669) (-1.2272727,-0.1807438) (-1.1727273,-0.1650347) (-1.1181818,-0.1500397)
    (-1.0636364,-0.1357587) (-1.0090909,-0.1221917) (-0.9545455,-0.1093388) (-0.9000000,-0.0972000) (-0.8454545,-0.0857752)
    (-0.7909091,-0.0750645) (-0.7363636,-0.0650678) (-0.6818182,-0.0557851) (-0.6272727,-0.0472165) (-0.5727273,-0.0393620)
    (-0.5181818,-0.0322215) (-0.4636364,-0.0257950) (-0.4090909,-0.0200826) (-0.3545455,-0.0150843) (-0.3000000,-0.0108000)
    (-0.2454545,-0.0072298) (-0.1909091,-0.0043736) (-0.1363636,-0.0022314) (-0.0818182,-0.0008033) (-0.0272727,-0.0000893)
    (0.0272727,-0.0000893) (0.0818182,-0.0008033) (0.1363636,-0.0022314) (0.1909091,-0.0043736) (0.2454545,-0.0072298)
    (0.3000000,-0.0108000) (0.3545455,-0.0150843) (0.4090909,-0.0200826) (0.4636364,-0.0257950) (0.5181818,-0.0322215)
    (0.5727273,-0.0393620) (0.6272727,-0.0472165) (0.6818182,-0.0557851) (0.7363636,-0.0650678) (0.7909091,-0.0750645)
    (0.8454545,-0.0857752) (0.9000000,-0.0972000) (0.9545455,-0.1093388) (1.0090909,-0.1221917) (1.0636364,-0.1357587)
    (1.1181818,-0.1500397) (1.1727273,-0.1650347) (1.2272727,-0.1807438) (1.2818182,-0.1971669) (1.3363636,-0.2143041)
    (1.3909091,-0.2321554) (1.4454545,-0.2507207) (1.5000000,-0.2700000) (1.5545455,-0.2899934) (1.6090909,-0.3107008)
    (1.6636364,-0.3321223) (1.7181818,-0.3542579) (1.7727273,-0.3771074) (1.8272727,-0.4006711) (1.8818182,-0.4249488)
    (1.9363636,-0.4499405) (1.9909091,-0.4756463) (2.0454545,-0.5020661) (2.1000000,-0.5292000) (2.1545455,-0.5570479)
    (2.2090909,-0.5856099) (2.2636364,-0.6148860) (2.3181818,-0.6448760) (2.3727273,-0.6755802) (2.4272727,-0.7069983)
    (2.4818182,-0.7391306) (2.5363636,-0.7719769) (2.5909091,-0.8055372) (2.6454545,-0.8398116) (2.7000000,-0.8748000)
  };
  \fill (P) circle (1.4pt) node[below right] {$P_0$};
  \coordinate (G0) at (0,1.8);
  \coordinate (Gt) at (1.55,1.35);
  \fill (G0) circle (1.4pt) node[above] {$G(0)$};
  \fill (Gt) circle (1.4pt) node[right] {$G(t)$};
  \draw[thick] (P) -- (G0) node[midway,left] {$l$};
  \draw[dashed] (P) -- (Gt) node[midway,below right] {$l'(t)\ge l$};
  \draw[->] (G0) -- (Gt) node[midway,above=5pt] {$\mathbf u(t)$};
  \node[align=center] at (0,-2.25) {the center of mass remains outside\\the initial exclusion sphere};
\end{tikzpicture}
\caption{Geometry of the center-of-mass barrier theorem.  The initial center
of mass lies on the sphere of radius $l$ centered at the fixed apex $P_0$.
If every admissible configuration keeps the center of mass on or outside this
sphere, the available gravitational height loss is at most quadratic in the
center-of-mass displacement. The displaced point illustrates the geometric
inequality only; it is not a trajectory starting from exact rest.}
\label{fig:com-barrier}
\end{figure}
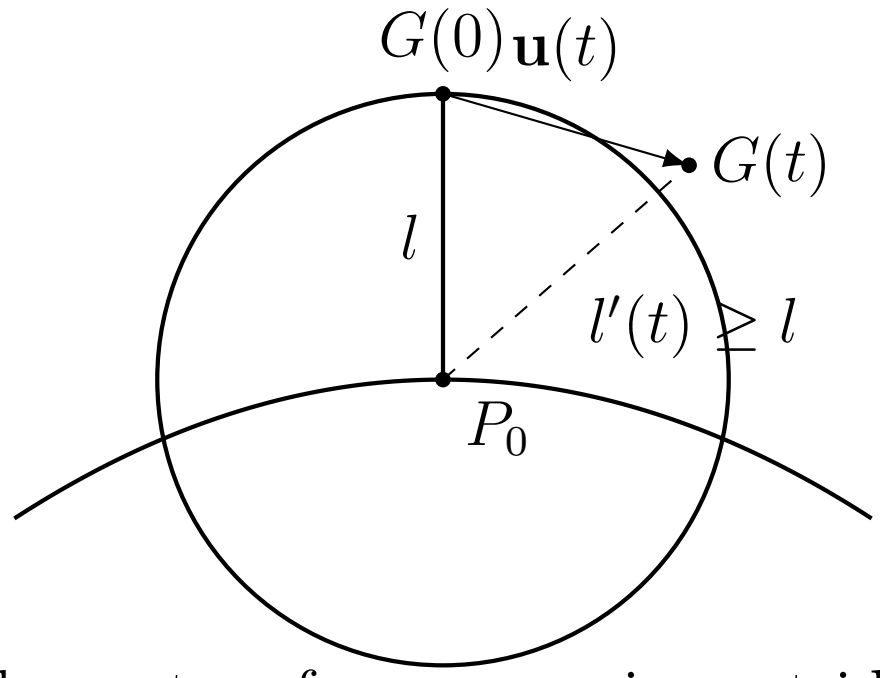

Figure~\ref{fig:com-barrier} illustrates the exclusion sphere in condition (B2) of Theorem~\ref{thm:com-barrier}.

\begin{theorem}[Center-of-mass barrier theorem]
\label{thm:com-barrier}
Assume that the material trajectories are absolutely continuous, that total
kinetic energy is nonnegative and positive definite in the material
velocities, and that the following conditions hold. Velocity inequalities
are understood almost everywhere in time; the energy bound holds throughout
the interval considered:
\begin{enumerate}
\item[(B1)] \textbf{Exact initial rest:} $K(0)=0$.
\item[(B2)] \textbf{Center-of-mass barrier:}
\begin{equation}
 |G(t)-P_0|\ge |G(0)-P_0|=l
 \label{eq:com-barrier-condition}
\end{equation}
for every time in the interval considered.
For a sphere whose center of mass is its geometric center,
$l=\rho$, and this condition follows from nonpenetration: a distance
$|G(t)-P_0|<\rho$ would place the fixed support point $P_0$ inside the sphere.
\item[(B3)] \textbf{Translational kinetic lower bound:}
\begin{equation}
 K(t)\ge\frac12m\norm{\dot G(t)}^2.
 \label{eq:com-kinetic-lower}
\end{equation}
\item[(B4)] \textbf{No available energy beyond gravity:}
\begin{equation}
 K(t)\le mg\,[G_z(0)-G_z(t)]=mgH(t).
 \label{eq:gravity-only-bound}
\end{equation}
\end{enumerate}
Then
\begin{equation}
 G(t)=G(0),
 \qquad K(t)=0
\end{equation}
for all such times.  In particular, the system cannot remain at rest for a
finite waiting time and then spontaneously depart.
\end{theorem}

\begin{proof}
Condition~\eqref{eq:com-barrier-condition} is
\begin{equation}
 \norm{l\eZ+\mathbf u(t)}^2\ge l^2.
\end{equation}
Expanding gives
\begin{equation}
 l^2+2lu_z+\norm{\mathbf u}^2\ge l^2,
\end{equation}
so the center-of-mass height loss satisfies the quadratic bound
\begin{equation}
 H(t)=-u_z(t)\le\frac{D(t)^2}{2l}.
 \label{eq:com-quadratic-bound}
\end{equation}
Using assumptions (B3) and (B4),
\begin{equation}
 \norm{\dot G(t)}^2
 \le\frac{2K(t)}{m}
 \le2gH(t)
 \le\frac{g}{l}D(t)^2.
 \label{eq:com-speed-bound}
\end{equation}
Since $D(t)=\norm{G(t)-G(0)}$ is absolutely continuous,
\begin{equation}
 |D'(t)|\le\norm{\dot G(t)}
 \le\sqrt{\frac{g}{l}}\,D(t)
 \label{eq:com-linear-bound}
\end{equation}
for almost every $t$. Since $D(0)=0$, Lemma~\ref{lem:gronwall} gives
\begin{equation}
 D(t)\equiv0.
\end{equation}
Thus $G(t)=G(0)$ and $H(t)=0$.  Equation~\eqref{eq:gravity-only-bound} then
implies $0\le K(t)\le0$, hence $K(t)=0$.
\end{proof}


\paragraph{A sufficient geometry for the barrier condition.}
Theorem~\ref{thm:com-barrier} assumes a barrier for the center of mass.
For a sphere centered at its center of mass, nonpenetration supplies this
barrier directly. For a nonspherical body, a separate geometric argument
is needed. The following lemma supplies one for planar convex rolling.
It describes admissible configurations; the theorem then excludes motion
from rest under the stated energy assumptions.

\begin{lemma}[Barrier under planar convex rolling]
\label{lem:geo-com-barrier}
Consider a rigid body rolling without slipping in a fixed vertical plane
on a fixed support, with continuous single-point contact near its initial
contact point $P_0=0$. Parametrize the support and the body's initial lower
boundary by arclength from that point as
\begin{align}
 \mathbf S(r)=(x_1(r),-h_1(r)),\qquad
 \mathbf B(r)=(x_2(r),h_2(r)),
\label{eq:sb}
\end{align}
where both curves start at the origin and
\begin{align}
 x_i'(r)=\cos\theta_i(r),\qquad
 h_i'(r)=\sin\theta_i(r)\quad(i=1,2).
 \label{eq:xh}
\end{align}
Assume that $\theta_1,\theta_2$ are continuous, nonnegative and
nondecreasing on $[0,r_*]$, with $\theta_i(0)=0$ and
$\theta_1(r)+\theta_2(r)<\pi/2$.
Consider the rolling branch on which the contacting material point is
$\mathbf B(r)$ and the support contact is $\mathbf S(r)$; their arclengths
agree by the no-slip condition.
For any body-fixed point initially at $R(0)=L\eZ$, $L>0$, its position
$R(r)$ satisfies
\begin{equation}
 |R(r)-P_0|\ge L\qquad(0\le r\le r_*).
 \label{eq:point-barrier}
\end{equation}
In particular, if the center of mass is initially on this vertical line,
condition (B2) holds along this rolling branch.
\end{lemma}

\begin{proof}
Write $\theta=\theta_1(r)+\theta_2(r)$. Tangent matching requires the
clockwise rotation
\[
 Q_\theta=\begin{pmatrix}\cos\theta&\sin\theta\\
                         -\sin\theta&\cos\theta\end{pmatrix}.
\]
The original material contact point moves to
$\mathbf p=\mathbf S(r)-Q_\theta\mathbf B(r)$ (using Eq \eqref{eq:sb}), and
\begin{align}
 R(r)=\mathbf p+\mathbf v,\qquad
 \mathbf v=Q_\theta(0,L)=L(\sin\theta,\cos\theta).
\end{align}
In the second line we use Eq \eqref{eq:xh}.
Thus $|\mathbf v|=L$ and direct multiplication gives
\begin{align}
 \frac{\mathbf p\cdot\mathbf v}{L}
 &=x_1(r)\sin\theta-h_1(r)\cos\theta-h_2(r)\\
 &=\int_0^r\left[
 \sin\bigl(\theta-\theta_1(s)\bigr)-\sin\theta_2(s)
 \right]ds.
\end{align}
For $0\le s\le r$, monotonicity of the $\theta_1, \theta_2$ implies
\[
 0\le\theta_2(s)
 \le\theta_1(r)-\theta_1(s)+\theta_2(r)
 =\theta-\theta_1(s)<\pi/2.
\]
Recall $\theta=\theta_1(r)+\theta_2(r)$
Since sine is increasing on this interval, the integrand is nonnegative.
Consequently,
\[
 |R(r)-P_0|^2
 =L^2+|\mathbf p|^2+2\mathbf p\cdot\mathbf v\ge L^2.
\]
Taking $R=G$ and $L=l$ proves (B2).
\end{proof}

The angle assumptions express local convexity of the two contact profiles
in the stated sign convention. They allow unbounded curvature at the
initial point, as on the standard Norton dome. The lemma does not cover
arbitrary concave profiles or general three-dimensional rolling.

\paragraph{Relaxed deformable body.}
Consider a deformable body with total kinetic energy $K(t)$ and recoverable
internal energy $U(t)$.  Suppose
\begin{equation}
 K(t)+U(t)+mgG_z(t)
 \le U(0)+mgG_z(0)
 \label{eq:deformable-passive}
\end{equation}
and the initial state contains no releasable internal energy in the sense that
\begin{equation}
 U(t)\ge U(0)
 \label{eq:no-internal-release}
\end{equation}
for the configurations considered.  If the center-of-mass barrier
\eqref{eq:com-barrier-condition} holds, then the body cannot spontaneously
translate, rotate, or deform from exact rest.

\begin{proof}
Equations~\eqref{eq:deformable-passive} and
\eqref{eq:no-internal-release} imply
\begin{equation}
 K(t)\le mg[G_z(0)-G_z(t)],
\end{equation}
which is assumption (B4).  For any system of particles or continuum body, the
total kinetic energy decomposes into center-of-mass translation plus
nonnegative motion relative to the center of mass, so assumption (B3) holds.
Theorem~\ref{thm:com-barrier} gives $K=0$.  Positive definiteness of the kinetic energy in the material velocities then
forces those velocities to vanish (almost everywhere with respect to mass
for a continuum body).
\end{proof}

\paragraph{Necessary alternatives for a clay-like body.}
If a deformable body departs from exact rest at the apex,
then at least one of the following must occur:
\begin{enumerate}
\item its center of mass enters the open sphere
$|G(t)-P_0|<l$ at some time;
\item recoverable internal energy is released below its initial value; or
\item energy is injected by an external agency or by an active contact.
\end{enumerate}
Thus deformation by itself does not evade the theorem; it must change the
center-of-mass geometry or the energy budget.

\paragraph{Scope of the barrier hypothesis.}
Theorem~\ref{thm:com-barrier} is independent of the detailed dome profile,
contact law, and rigidity of the body.  All model-specific geometry has been
compressed into the single condition $|G(t)-P_0|\ge l$.  Proving that condition
for a particular rolling or deforming body is a separate geometric problem.
In particular, positive volume or a no-slip condition alone is not a
verification of this hypothesis. For a nonspherical rolling body, the
contact geometry and the changing orientation must both be accounted for.
The next section supplies a sufficient criterion without requiring a
separate rolling-trajectory construction.

\section{The interior-tangent-ball criterion for rigid bodies}
\label{sec:rigid}

An interior tangent ball gives a different geometric criterion for rigid
bodies, without requiring planar motion or rolling without slipping.
Nonpenetration constrains the center of this auxiliary ball, which need not
coincide with the center of mass. The center-of-mass barrier of
Theorem~\ref{thm:com-barrier} therefore does not follow automatically.
Instead, we control height loss by both translation of the ball center
and rotation of the body. Two lemmas separate this geometric estimate
from the kinetic estimate used in the final theorem.

Let a fixed support occupy a closed region $\Ocal\subset\R^3$, and let
$P\in\partial\Ocal$ be the initial contact point.  Choose coordinates such
that
\begin{equation}
 P=0,
 \qquad
 \eZ=(0,0,1),
\end{equation}
and gravity is $-g\eZ$.  Let $\Bcal_0\subset\R^3$ be the initial configuration
of a rigid body of mass $m>0$.  A rigid motion is
\begin{equation}
 \Bcal(t)=\mathbf x(t)+Q(t)\Bcal_0,
 \qquad Q(t)\in SO(3),
\end{equation}
with $\mathbf x(0)=0$ and $Q(0)=I$. Assume that $\mathbf x$ and $Q$
are absolutely continuous. Let $G(t)$ be the center of mass and
$\bm\omega(t)$ the angular velocity, defined almost everywhere.
Differential and kinetic identities below are understood almost everywhere;
the energy inequality is imposed throughout the interval, with prescribed
initial kinetic energy $K(0)=0$.

We impose the following assumptions.

\begin{enumerate}
\item[(A1)] \textbf{Interior tangent ball.}
There exists $b>0$ with center $A_0=b\eZ$ such that
\begin{equation}
 \overline{B_b(A_0)}\subset\Bcal_0,
 \qquad
 P\in\partial B_b(A_0)\cap\partial\Bcal_0.
 \label{eq:interior-ball-condition}
\end{equation}

\item[(A2)] \textbf{Center-of-mass alignment.}
For some $d\in\R$,
\begin{equation}
 G_0=A_0+d\eZ.
\end{equation}

\item[(A3)] \textbf{Positive inertia.}
The inertia tensor about the center of mass is positive definite, and its
smallest principal moment is denoted $J_{\min}>0$.

\item[(A4)] \textbf{Nonpenetration.}
The moving body and the fixed support do not interpenetrate; in particular,
the fixed point $P$ never lies in the interior of the moving body:
$P\notin\operatorname{int}\Bcal(t)$.

\item[(A5)] \textbf{Passive contact.}
The contact does not inject mechanical energy:
\begin{equation}
 K(t)+mgG_z(t)
 \le K(0)+mgG_z(0).
 \label{eq:passive-energy}
\end{equation}

\item[(A6)] \textbf{Exact initial rest.}
\begin{equation}
 \dot G(0)=0,
 \qquad
 \bm\omega(0)=0.
\end{equation}
\end{enumerate}

\begin{lemma}[Geometric height bound]
\label{lem:rigid-height}
Under (A1), (A2), and (A4), let $A(t)=\mathbf x(t)+Q(t)A_0$,
$\mathbf u=A-A_0$, $\mathbf n=Q\eZ$, and
$\mathbf y=(\mathbf u,b(\mathbf n-\eZ))$.
The center-of-mass height loss $H=G_z(0)-G_z(t)$ satisfies
\[
 H\le C_H\norm{\mathbf y}^2,\qquad
 C_H=\max\{1/(2b),|d|/(2b^2)\}.
\]
\end{lemma}
\begin{proof}

The interior ball is a geometric marker rigidly embedded in the body.  Its
center moves as
\begin{equation}
 A(t)=\mathbf x(t)+Q(t)A_0.
\end{equation}
Define
\begin{equation}
 \mathbf u(t)=A(t)-A_0,
 \qquad
 \mathbf n(t)=Q(t)\eZ.
\end{equation}
Since the fixed point $P=0$ belongs to the support, nonpenetration implies that
$P$ cannot enter the moving interior ball.  Hence
\begin{equation}
 \norm{A(t)}\ge b.
\end{equation}
Using $A=b\eZ+\mathbf u$,
\begin{equation}
 \norm{b\eZ+\mathbf u}^2\ge b^2,
\end{equation}
and therefore
\begin{equation}
 -u_z\le\frac{\norm{\mathbf u}^2}{2b}.
 \label{eq:basic-interior-bound}
\end{equation}
This is the key geometric estimate.  It is the elementary finite-dimensional
form of the quadratic inequalities associated with interior-sphere/proximal
geometry; see Ref.~\cite{ClarkeSternWolenski1995}.

Because rigid motion preserves the relative vector between $A$ and the center
of mass,
\begin{equation}
 G(t)=A(t)+d\mathbf n(t).
\end{equation}
Let
\begin{equation}
 H(t)=G_z(0)-G_z(t)
\end{equation}
be the center-of-mass downward displacement.  Then
\begin{equation}
 H=-u_z+d(1-n_z).
\end{equation}
Since $\mathbf n$ and $\eZ$ are unit vectors,
\begin{equation}
 1-n_z=\frac12\norm{\mathbf n-\eZ}^2.
\end{equation}
Thus Eq.~\eqref{eq:basic-interior-bound} yields
\begin{equation}
 H(t)
 \le\frac{\norm{\mathbf u(t)}^2}{2b}
 +\frac{|d|}{2}\norm{\mathbf n(t)-\eZ}^2.
 \label{eq:H-bound}
\end{equation}
Define
\begin{equation}
 \mathbf y(t)=
 \bigl(\mathbf u(t),\,b[\mathbf n(t)-\eZ]\bigr)\in\R^6.
\end{equation}
Then
\begin{equation}
 H(t)\le C_H\norm{\mathbf y(t)}^2,
 \label{eq:H-quadratic}
\end{equation}
where one may take
\begin{equation}
 C_H=\max\left\{\frac1{2b},\frac{|d|}{2b^2}\right\}.
\end{equation}

\end{proof}

\begin{lemma}[Kinetic control]
\label{lem:rigid-kinetic}
Under (A2) and (A3), the displacement $\mathbf y$ defined in
Lemma~\ref{lem:rigid-height} satisfies
\[
 \norm{\dot{\mathbf y}}^2\le C_KK,\qquad
 C_K=\max\{4/m,\,2(2d^2+b^2)/J_{\min}\}.
\]
\end{lemma}
\begin{proof}

The rigid-body kinetic energy is
\begin{equation}
 K(t)=\frac12m\norm{\dot G(t)}^2
 +\frac12\bm\omega(t)^{\mathsf T}J(t)\bm\omega(t).
\end{equation}
Therefore
\begin{equation}
 K(t)\ge
 \frac12m\norm{\dot G(t)}^2
 +\frac12J_{\min}\norm{\bm\omega(t)}^2.
 \label{eq:Klower}
\end{equation}
Rigid-body kinematics gives
\begin{equation}
 \dot{\mathbf n}=\bm\omega\times\mathbf n,
 \qquad
 \norm{\dot{\mathbf n}}\le\norm{\bm\omega}.
\end{equation}
Since $G=A+d\mathbf n$,
\begin{equation}
 \dot{\mathbf u}=\dot G-d\dot{\mathbf n}.
\end{equation}
Hence
\begin{align}
 \norm{\dot{\mathbf y}}^2
 &=\norm{\dot{\mathbf u}}^2
 +b^2\norm{\dot{\mathbf n}}^2\\
 &\le2\norm{\dot G}^2
 +(2d^2+b^2)\norm{\bm\omega}^2.
\end{align}
Combining with Eq.~\eqref{eq:Klower},
\begin{equation}
 \norm{\dot{\mathbf y}}^2\le C_KK,
 \label{eq:ydotK}
\end{equation}
where, for example,
\begin{equation}
 C_K=\max\left\{
 \frac4m,
 \frac{2(2d^2+b^2)}{J_{\min}}
 \right\}.
\end{equation}

\end{proof}

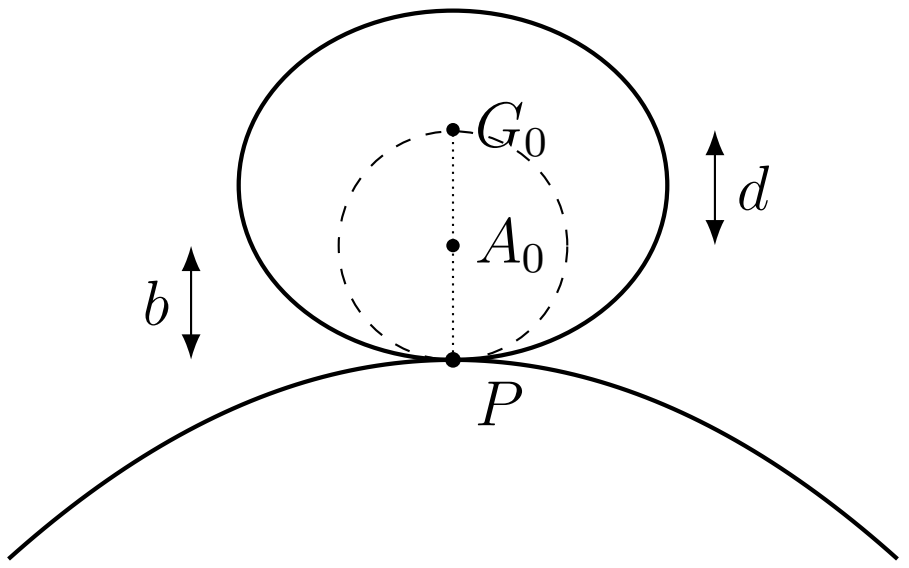
\begin{figure}[H]
\centering
\begin{tikzpicture}[scale=1.05,>=Latex]
  \draw[thick] plot coordinates {
    (-2.8000000,-1.2544000) (-2.7434343,-1.2042291) (-2.6868687,-1.1550821) (-2.6303030,-1.1069590) (-2.5737374,-1.0598599)
    (-2.5171717,-1.0137846) (-2.4606061,-0.9687331) (-2.4040404,-0.9247056) (-2.3474747,-0.8817020) (-2.2909091,-0.8397223)
    (-2.2343434,-0.7987665) (-2.1777778,-0.7588346) (-2.1212121,-0.7199265) (-2.0646465,-0.6820424) (-2.0080808,-0.6451822)
    (-1.9515152,-0.6093458) (-1.8949495,-0.5745334) (-1.8383838,-0.5407448) (-1.7818182,-0.5079802) (-1.7252525,-0.4762394)
    (-1.6686869,-0.4455225) (-1.6121212,-0.4158296) (-1.5555556,-0.3871605) (-1.4989899,-0.3595153) (-1.4424242,-0.3328940)
    (-1.3858586,-0.3072966) (-1.3292929,-0.2827232) (-1.2727273,-0.2591736) (-1.2161616,-0.2366479) (-1.1595960,-0.2151460)
    (-1.1030303,-0.1946681) (-1.0464646,-0.1752141) (-0.9898990,-0.1567840) (-0.9333333,-0.1393778) (-0.8767677,-0.1229954)
    (-0.8202020,-0.1076370) (-0.7636364,-0.0933025) (-0.7070707,-0.0799918) (-0.6505051,-0.0677051) (-0.5939394,-0.0564422)
    (-0.5373737,-0.0462033) (-0.4808081,-0.0369882) (-0.4242424,-0.0287971) (-0.3676768,-0.0216298) (-0.3111111,-0.0154864)
    (-0.2545455,-0.0103669) (-0.1979798,-0.0062714) (-0.1414141,-0.0031997) (-0.0848485,-0.0011519) (-0.0282828,-0.0001280)
    (0.0282828,-0.0001280) (0.0848485,-0.0011519) (0.1414141,-0.0031997) (0.1979798,-0.0062714) (0.2545455,-0.0103669)
    (0.3111111,-0.0154864) (0.3676768,-0.0216298) (0.4242424,-0.0287971) (0.4808081,-0.0369882) (0.5373737,-0.0462033)
    (0.5939394,-0.0564422) (0.6505051,-0.0677051) (0.7070707,-0.0799918) (0.7636364,-0.0933025) (0.8202020,-0.1076370)
    (0.8767677,-0.1229954) (0.9333333,-0.1393778) (0.9898990,-0.1567840) (1.0464646,-0.1752141) (1.1030303,-0.1946681)
    (1.1595960,-0.2151460) (1.2161616,-0.2366479) (1.2727273,-0.2591736) (1.3292929,-0.2827232) (1.3858586,-0.3072966)
    (1.4424242,-0.3328940) (1.4989899,-0.3595153) (1.5555556,-0.3871605) (1.6121212,-0.4158296) (1.6686869,-0.4455225)
    (1.7252525,-0.4762394) (1.7818182,-0.5079802) (1.8383838,-0.5407448) (1.8949495,-0.5745334) (1.9515152,-0.6093458)
    (2.0080808,-0.6451822) (2.0646465,-0.6820424) (2.1212121,-0.7199265) (2.1777778,-0.7588346) (2.2343434,-0.7987665)
    (2.2909091,-0.8397223) (2.3474747,-0.8817020) (2.4040404,-0.9247056) (2.4606061,-0.9687331) (2.5171717,-1.0137846)
    (2.5737374,-1.0598599) (2.6303030,-1.1069590) (2.6868687,-1.1550821) (2.7434343,-1.2042291) (2.8000000,-1.2544000)
  };
  \fill (0,0) circle (1.4pt) node[below right] {$P$};
  \draw[thick] (0,1.10) ellipse [x radius=1.35,y radius=1.10];
  \draw[densely dotted] (0,0) -- (0,1.45);
  \draw[dashed] (0,0.72) circle (0.72);
  \fill (0,0.72) circle (1.2pt) node[right] {$A_0$};
  \fill (0,1.45) circle (1.2pt) node[right] {$G_0$};
  \draw[<->] (-1.65,0) -- (-1.65,0.72)
    node[midway,left] {$b$};
  \draw[<->] (1.65,0.72) -- (1.65,1.45)
    node[midway,right] {$d$};
\end{tikzpicture}
\caption{A general rigid body containing an interior tangent ball of radius
$b$ at the initial contact point.  The auxiliary ball is carried rigidly with
the body and supplies the quadratic geometric estimate used in
Theorem~\ref{thm:interior-ball}.}
\label{fig:interior-ball}
\end{figure}

Figure~\ref{fig:interior-ball} shows the interior tangent ball and center-of-mass alignment assumed in Theorem~\ref{thm:interior-ball}.

\begin{theorem}[Interior-tangent-ball criterion]
\label{thm:interior-ball}
Under assumptions (A1)--(A6), any absolutely continuous admissible rigid-body
motion starting from the prescribed initial rest state is stationary.  In
particular, the body cannot remain at rest for an arbitrary finite waiting time
and then spontaneously depart.
\end{theorem}

\begin{proof}
Initial rest and passive contact give
\begin{equation}
 0\le K(t)\le mgH(t).
 \label{eq:energy-to-height}
\end{equation}
Lemmas~\ref{lem:rigid-height} and \ref{lem:rigid-kinetic} imply
\begin{equation}
 \norm{\dot{\mathbf y}}\le L\norm{\mathbf y},
 \qquad L=\sqrt{C_KmgC_H}.
 \label{eq:linear-y}
\end{equation}
Since $\mathbf y$ is absolutely continuous and $\mathbf y(0)=0$,
\[
 \norm{\mathbf y(t)}\le L\int_0^t\norm{\mathbf y(s)}\,ds.
\]
Lemma~\ref{lem:gronwall}, applied to $D(t)=\norm{\mathbf y(t)}$, gives
$\mathbf y=0$. Thus $H=0$ and
Eq.~\eqref{eq:energy-to-height} gives $K=0$.
Positive mass and positive-definite inertia then imply
$\dot G=0$ and $\bm\omega=0$ almost everywhere. Absolute continuity
ensures that the full rigid-body configuration remains fixed.
\end{proof}

\paragraph{Finite sphere as a special case.}
A rigid sphere of radius $\rho>0$ with its center of mass at its geometric
center, satisfying the same passive-contact and
initial-rest assumptions obeys Theorem~\ref{thm:interior-ball}: choose the
sphere itself as the interior tangent ball, so that $b=\rho$ and $d=0$.

\paragraph{Nonzero volume is not the theorem.}
Positive volume alone does not imply an interior tangent ball at a specified
contact point.  A body may, for example, have a sharp tip.  The theorem is
therefore intentionally stated as a sufficient local finite-thickness
criterion and makes no claim for arbitrary positive-volume bodies.

\section{Summary of the geometric uniqueness mechanism}
\label{sec:essence}
\label{sec:osgood}

The preceding results progress from an explicit sphere calculation to two
geometric sufficient conditions for more general bodies. Their common
mechanism is a bound on the gravitational height available to drive motion. Let $D\ge0$ measure displacement from the initial configuration,
$H$ denote center-of-mass height loss, and $K$ total kinetic energy. Geometry
controls $H$ in terms of $D$; a kinetic estimate and passivity then control
the rate at which $D$ can grow. The following standard energy argument
summarizes the estimates established in the preceding sections.

\begin{lemma}[Quadratic barrier principle]
\label{prop:critical-exponent}
Let $D:[0,T]\to[0,\infty)$ be absolutely continuous with $D(0)=0$.
Suppose that, while $D$ is sufficiently small,
\begin{equation}
 H\le C_HD^2,\qquad |D'|^2\le C_KK,\qquad 0\le K\le mgH,
 \label{eq:meta-bounds}
\end{equation}
where $m,g,C_H,C_K$ are positive finite constants and differential
inequalities hold almost everywhere. Then $D(t)=0$ throughout $[0,T]$.
\end{lemma}

\begin{proof}
With $L=(C_KmgC_H)^{1/2}$, the assumptions give $|D'|\le LD$.
Absolute continuity and $D(0)=0$ imply
$D(t)\le L\int_0^t D(s)\,ds$, so Lemma~\ref{lem:gronwall} gives $D=0$.
If the bounds are local, apply this argument up to the first exit from their
neighborhood; continuity excludes such an exit.
\end{proof}

Thus the shared structure is
\begin{equation}
\begin{gathered}
 \text{geometric barrier}\ \Longrightarrow\ H\le C_HD^2,\\
 |D'|^2\le C_KK\le C_KmgH
 \ \Longrightarrow\ |D'|\le LD
 \ \Longrightarrow\ D(t)\equiv0.
 \end{gathered}
 \label{eq:essence-chain}
\end{equation}
The required geometry is established explicitly for a sphere, assumed as a
center-of-mass exclusion condition for a passive system, or derived from an
interior tangent ball for an aligned rigid body. These are distinct sufficient
conditions, not interchangeable descriptions of all extended bodies.

Table~\ref{tab:essence} summarizes the progression of height and speed laws.
The last two rows are distinct sufficient criteria, not successive
assumptions implying one another. Here
$r$ is contact arclength, $q$ sphere-center arclength, $\rho>0$ sphere radius,
$l>0$ the initial apex-to-center distance, and $\mathbf y$ the rigid-body
displacement defined in Sec.~\ref{sec:rigid}; in the last row
$Y=\|\mathbf y\|$. The exponent $a$ parametrizes the family
$\ddot r=cr^{1/a}$ introduced in Sec.~\ref{sec:generalized}.

\begin{table}[t]
\centering
\renewcommand{\arraystretch}{1.25}
\begin{tabular}{>{\raggedright\arraybackslash}p{0.25\textwidth}
                >{\centering\arraybackslash}p{0.20\textwidth}
                >{\centering\arraybackslash}p{0.20\textwidth}
                >{\centering\arraybackslash}p{0.20\textwidth}}
\toprule
Model & Height law & Speed law & Escape integral \\
\midrule
Standard Norton point particle
& $H\propto r^{3/2}$
& $|\dot r|\propto r^{3/4}$
& $\int_0 dr/r^{3/4}<\infty$ \\
Finite sphere, $a>1$
& $H\sim q^2/(2\rho)$
& $|\dot q|\sim\sqrt{g/\rho}\,q$
& $\int_0 dq/q=\infty$ \\
Center-of-mass barrier
& $H\le D^2/(2l)$
& $|\dot D|\le\sqrt{g/l}\,D$
& $\int_0 dD/D=\infty$ \\
Interior tangent ball
& $H\le C_H\|\mathbf y\|^2$
& $\|\dot{\mathbf y}\|\le L\|\mathbf y\|$
& $\int_0 dY/Y=\infty$ \\
\bottomrule
\end{tabular}
\caption{The same integrability threshold appears at every level of the
analysis.  The point-particle Norton model is subcritical; the geometries satisfying
our sufficient conditions impose a quadratic upper bound on height loss.}
\label{tab:essence}
\end{table}

The travel-time calculations can also be expressed through the following
integral criterion. It also explains why a quadratic height bound is the threshold
for this particular energy argument.

\begin{lemma}[Escape-time criterion]
\label{lem:escape}
Let $D:[0,T]\to[0,R]$ be absolutely continuous with $D(0)=0$.  Suppose
\begin{equation}
 |D'(t)|\le\Phi(D(t))
 \label{eq:escape-bound}
\end{equation}
for almost every $t$, where $\Phi:[0,R]\to[0,\infty)$ is continuous and
$\Phi(s)>0$ for $s>0$.  If
\begin{equation}
 \int_{0^+}\frac{ds}{\Phi(s)}=+\infty,
 \label{eq:osgood-integral}
\end{equation}
then $D(t)\equiv0$ on $[0,T]$.
\end{lemma}

\begin{proof}
Suppose that some positive level $\varepsilon$ is reached at a first time
$t_\varepsilon$.  For $0<\delta<\varepsilon$, let $t_\delta$ be the last time
before $t_\varepsilon$ at which $D=\delta$.  On
$[t_\delta,t_\varepsilon]$, the chain rule for absolutely continuous functions
and Eq.~\eqref{eq:escape-bound} give
\begin{align}
 \int_\delta^\varepsilon\frac{ds}{\Phi(s)}
 &=\int_{t_\delta}^{t_\varepsilon}
 \frac{D'(t)}{\Phi(D(t))}\,dt\\
 &\le\int_{t_\delta}^{t_\varepsilon}
 \frac{|D'(t)|}{\Phi(D(t))}\,dt
 \le t_\varepsilon-t_\delta
 \le T.
\end{align}
Letting $\delta\to0^+$ contradicts
Eq.~\eqref{eq:osgood-integral}.  Thus no positive level can be reached.
\end{proof}

This is the relevant Osgood-type uniqueness mechanism \cite{Osgood1898}.
The zero-initial-data conclusion of Lemma~\ref{lem:gronwall} is the
linear case $\Phi(D)=LD$ with $L>0$.

More generally, a bound $H\le C_HD^p$ with the same kinetic and energy
estimates gives $|D'|\le CD^{p/2}$. The reciprocal-speed integral diverges
for $p\ge2$, excluding departure from zero; $p=2$ is the logarithmic case.
For $p<2$, this upper bound alone does not establish nonuniqueness. The
point-particle models in Sec.~\ref{sec:generalized} establish it through explicit departing solutions.

\section{Discussion and conclusions}
\label{sec:discussion}

The results preserve Norton's point-particle construction while showing that
its uniqueness structure changes under a specific geometric modification.
For $\ddot r=cr^{1/a}$, delayed departure from exact rest occurs precisely for
$a>1$. Every fixed positive sphere radius excludes such departure for all
finite $a>0$. In the nonunique point-particle regime, the universal center
geometry $\Delta h\sim q^2/(2\rho)$ explains the logarithmically divergent
escape time, while Eq.~\eqref{eq:lipschitz-force} gives a direct acceleration
bound. Rotational inertia changes the time scale but is not the source of
this regularization.

The zero-radius limit is singular: the coefficient $\sqrt{g/\rho}$ in the
near-apex speed law diverges as $\rho\to0^+$. Likewise, the constants in the
center-of-mass and interior-ball estimates diverge as $l\to0^+$ or $b\to0^+$.
Consequently, a uniqueness estimate for fixed positive size cannot simply be
evaluated at zero size. Nor does uniqueness of exact rest imply stability:
small displacements can grow even though spontaneous departure is excluded.

The two general criteria separate the geometry from the dynamics. The
center-of-mass theorem allows deformation, but its exclusion condition must
be established for the body in question and no internal energy may be
released. The rigid-body theorem obtains quadratic height control from an
interior tangent ball and vertical center-of-mass alignment. Neither theorem
asserts that positive volume alone is sufficient. A sharp contact may fail
the interior-ball condition, and a deformable body may violate the
center-of-mass barrier. These explicit hypotheses distinguish the present
results from a general claim about determinism in classical mechanics.

The analysis connects standard uniqueness and contact-geometry tools
\cite{Osgood1898,ClarkeSternWolenski1995,Ballard2000} to an explicit mechanics
example. Its contribution is the finite-size classification and the geometric
explanation of the critical height law, rather than a new general uniqueness
theory.

For advanced undergraduate mechanics, the sphere calculation provides a
concrete example of how a geometric idealization changes uniqueness of
motion. The subsequent energy estimates connect this example to differential
inequalities and distinguish uniqueness from stability.

\section*{Acknowledgements}
EJK acknowledges financial support from the National Science and Technology
Council (NSTC) of Taiwan under Grant No.~NSTC~114-2112-M-A49-036-MY3.

\medskip
\noindent\textbf{Author contributions.}
Guan-Chuan Chen conceived the project and carried out the explicit mechanical
analysis of finite spheres on the standard and generalized Norton domes.
En-Jui Kuo identified Gr\"onwall's inequality as the unifying principle behind
the no-departure arguments and derived the interior-tangent-ball criterion.
Building on Chen's initial geometric arguments, both authors jointly developed
the abstract center-of-mass barrier formulation and its extension incorporating
internal energy, and completed the manuscript through discussion and revision.

\medskip
\noindent\textbf{Competing interests.}
The authors declare no competing interests.

\medskip
\noindent\textbf{AI disclosure.}
During preparation of this manuscript, we used OpenAI Codex to edit the
text, organize the manuscript, and assist in checking and revising
mathematical arguments, including the assumptions and proof of the planar
rolling lemma. All authors take full responsibility for the work presented
here.

\medskip
\noindent\textbf{Data availability statement}
This is an analytical study. No experimental or simulation datasets are reported.

\bibliographystyle{unsrt}
\bibliography{ref}

\end{document}